\documentclass[aps,prl,twocolumn,showpacs,amsmath,amssymb,superscriptaddress]{revtex4}
\pdfoutput=1
\usepackage{graphicx}
\usepackage{amssymb}
\usepackage{natbib}
\usepackage{calc}
\usepackage{SIunits}
\usepackage{textcomp}

\begin{document}

\title{Odd and Even Modes of Neutron Spin Resonance in the Bilayer Iron-Based Superconductor CaKFe$_4$As$_4$}

\author{Tao Xie}
\affiliation{Beijing National Laboratory for Condensed Matter
Physics, Institute of Physics, Chinese Academy of Sciences, Beijing
100190, China}
\affiliation{University of Chinese Academy of Sciences, Beijing 100049, China}
\author{Yuan Wei}
\affiliation{Beijing National Laboratory for Condensed Matter
Physics, Institute of Physics, Chinese Academy of Sciences, Beijing
100190, China}
\affiliation{University of Chinese Academy of Sciences, Beijing 100049, China}
\author{Dongliang Gong}
\affiliation{Beijing National Laboratory for Condensed Matter
Physics, Institute of Physics, Chinese Academy of Sciences, Beijing
100190, China}
\affiliation{University of Chinese Academy of Sciences, Beijing 100049, China}
\author{Tom Fennell}
\affiliation{Laboratory for Neutron Scattering and Imaging, Paul Scherrer Institute, CH-5232 Villigen, Switzerland}
\author{Uwe Stuhr}
\affiliation{Laboratory for Neutron Scattering and Imaging, Paul Scherrer Institute, CH-5232 Villigen, Switzerland}
\author{Ryoichi Kajimoto}
\affiliation{Materials and Life Science Division, J-PARC Center, Japan Atomic Energy Agency, Tokai, Ibaraki 319-1195, Japan}
\author{Kazuhiko Ikeuchi}
\affiliation{Neutron Science and Technology Center, Comprehensive Research Organization for Science and Society,
Tokai, Ibaraki 319-1106, Japan}
\author{Shiliang Li}
\affiliation{Beijing National Laboratory for Condensed Matter
Physics, Institute of Physics, Chinese Academy of Sciences, Beijing
100190, China}
\affiliation{University of Chinese Academy of Sciences, Beijing 100049, China}
\affiliation{Collaborative Innovation Center of Quantum Matter, Beijing 100190, China}
\author{Jiangping Hu}
\affiliation{Beijing National Laboratory for Condensed Matter
Physics, Institute of Physics, Chinese Academy of Sciences, Beijing
100190, China}
\affiliation{University of Chinese Academy of Sciences, Beijing 100049, China}
\affiliation{Collaborative Innovation Center of Quantum Matter, Beijing 100190, China}
\author{Huiqian Luo}
\email{hqluo@iphy.ac.cn}
\affiliation{Beijing National Laboratory for Condensed Matter
Physics, Institute of Physics, Chinese Academy of Sciences, Beijing
100190, China}

\date{\today}

\begin{abstract}
We report an inelastic neutron scattering study on the spin resonance in the bilayer iron-based superconductor CaKFe$_4$As$_4$. In contrast to its quasi-two-dimensional electron structure, three strongly $L$-dependent modes of spin resonance are found below $T_c=35$ K. The mode energies are below and linearly scale with the total superconducting gaps summed on the nesting hole and electron pockets, essentially in agreement with the results in cuprate and heavy fermion superconductors. This observation supports the sign-reversed Cooper pairing mechanism under multiple pairing channels and resolves the long-standing puzzles concerning the broadening and dispersive spin resonance peak in iron pnictides. More importantly, the triple resonant modes can be classified into odd and even symmetries with respect to the distance of Fe-Fe planes within the Fe-As bilayer unit. Thus, our results closely resemble those in the bilayer cuprates with nondegenerate spin excitations, suggesting that these two high-$T_c$ superconducting families share a common nature.

\end{abstract}

\maketitle

Understanding the superconducting mechanism in unconventional superconductors such as copper oxides, heavy fermions, iron pnictides, and iron chalcogenides, is one of the most important topics in modern condensed matter physics \cite{jttranquada2014,inosov2016,pdai2015}. On cooling below the superconducting transition temperature ($T_c$) in these materials, the spin excitations form a resonant peak with enhanced susceptibility at a certain energy and around the antiferromagnetic (AFM) wavevector of the parent compounds. This so-called neutron spin resonance, which is argued to be a spin-1 collective mode of particle-hole excitations in the superconducting state, gives prominent evidence for the magnetic Cooper pairing in cuprates and heavy fermion superconductors \cite{pdai2015,ostockert2011,meschrig2006,ysidis2007,gyu2009}.

The multiband physics from Fe $3d$ orbitals in iron-based superconductors opens a new opportunity to explore the unconventional superconductivity \cite{si2016,prichard2011}. In iron pnictides or chalcogenides, the sign-reversed $s$-wave ($s_\pm$) Cooper pairing can be obtained in both weak coupling approaches \cite{korshunov2008,avchubukov2008,tamaier2009,mazin2009} and strong correlated electron models \cite{Seo2008} and has been supported by many experimental evidences \cite{thanaguri2010,hyang2013,aakalenyuk2017,zydu2016,zydu2018}. In the $s_\pm$ superconducting state, a spin resonance is theoretically predicted to arise at the wave vector $\textbf{Q}$ linking between hole-electron or electron-electron pockets, which is experimentally observed in many systems \cite{christianson,yqiu2009,inosov2010,qureshi2012,wakimoto,zhang2013,pdjohnson,mywang2010,hqluo2013b,schi2009,zhang2011,chlee2013,jzhao2013,mgkim2013,zhang2013b,jtpark2011,gfriemel2012,masurmach2015,qswang2016a,mwma2017,txie2018}. If the resonance is indeed a spin exciton in the superconducting state, it should be a sharp peak bound at an energy ($E_R$) below the pair breaking energy, namely, the total superconducting gap summed on the two pockets linked by $\textbf{Q}$: $\Delta_{\rm tot}=|\Delta_k|+|\Delta_{k+Q}|$ \cite{avchubukov2008,tamaier2009,zhang2013b,lwharriger2012}. Here, $\Delta_k$ and $\Delta_{k+Q}$ have opposite signs (probably different values) to yield a finite coherence factor of this process, enhanced by the interband Fermi surface nesting under intraorbital Coulomb repulsion \cite{tamaier2009,mazin2009}. In contrast, a nonresonant broad peak in the magnetic spectrum below $T_c$ is expected above twice the superconducting gap ($2\Delta$) in the sign-preserved ($s_{++}$) state \cite{sonari2010,sonari2012,qswang2016}.  However, compared to the resonance mode observed in copper oxides \cite{meschrig2006,ysidis2007}, spin resonances in iron pnictides are actually much broader in energy distribution and more dispersive both in plane and along the $L$ direction \cite{schi2009,zhang2011,chlee2013,jzhao2013,mgkim2013}.  Lacking of the sharpness in both energy and momentum spaces may be attributed to the complex multi-orbital nature in iron-based superconductors that can lead to multiple resonant modes and spin anisotropy \cite{hqluo2013,steffens2013,mwma2017b,dhu2017,wwang2017}.

To further understand the spin resonance in iron-based superconductors, it is essential to make a full comparison to all behaviors of the resonant mode observed in cuprates.  In the hole-doped bilayer YBa$_2$Cu$_3$O$_{6+\delta}$ (YBCO) system, the spin resonance exhibits distinguished odd and even $L$ symmetries due to the nondegenerate interlayer magnetic excitations \cite{meschrig2006,spaihes2003,spaihes2004}, which is later confirmed in another bilayer system Bi$_2$Sr$_2$CaCu$_2$O$_{8+\delta}$ (Bi2212) \cite{ysidis2007,lcapogna2007}. These two different modes of spin resonance evolve in a strikingly similar doping dependence in both systems, and their separation in energy is fully determined by a weak AFM interaction between Cu-O planes within the bilayer unit, giving strong evidence for the magnetically mediated superconducting pairing mechanisms. However, this even mode has never been observed in iron-based superconductors, which seems to suggest that the spin resonance may have different origins in these two high-$T_c$ families.

In this Letter, we report an inelastic neutron scattering study on the spin excitations of stoichiometric iron-based superconductor CaKFe$_4$As$_4$ (1144 compound) with Fe-As bilayer structure [Fig. 1(a)]. Three spin resonance modes are identified at the wave vector ${\bf Q}$ from $\Gamma$ to $M$ point [Fig. 1(b)], where the resonance energies and the mode intensities are directly proportional to the total superconducting gaps summed on the nesting electron and hole bands. In contrast to its quasi-two-dimensional (2D) electron structure, the resonance intensity for all three modes is highly $L$ dependent with two opposite harmonic modulations, showing either odd symmetry $\sim\mid F(Q)\mid^2\sin^2(z\pi L)$ or even symmetry $\sim \mid F(Q)\mid^2\cos^2(z\pi L)$ [Figs. 1(c)$-$1(f)]. Here, $F(Q)$ is the magnetic form factor of $\mathrm{Fe}^{2+}$, and $zc=5.855$ \AA\ ($z=$ 0.4636) is the distance between adjacent Fe-Fe planes within the Fe-As bilayer unit [Fig. 1(a)]. We argue that such phenomenon is essentially similar to the nondegenerate bilayer magnetic excitations in YBCO \cite{meschrig2006} but under multiband pairing mechanism \cite{si2016}.

\begin{figure}[t]
\includegraphics[width=0.42\textwidth]{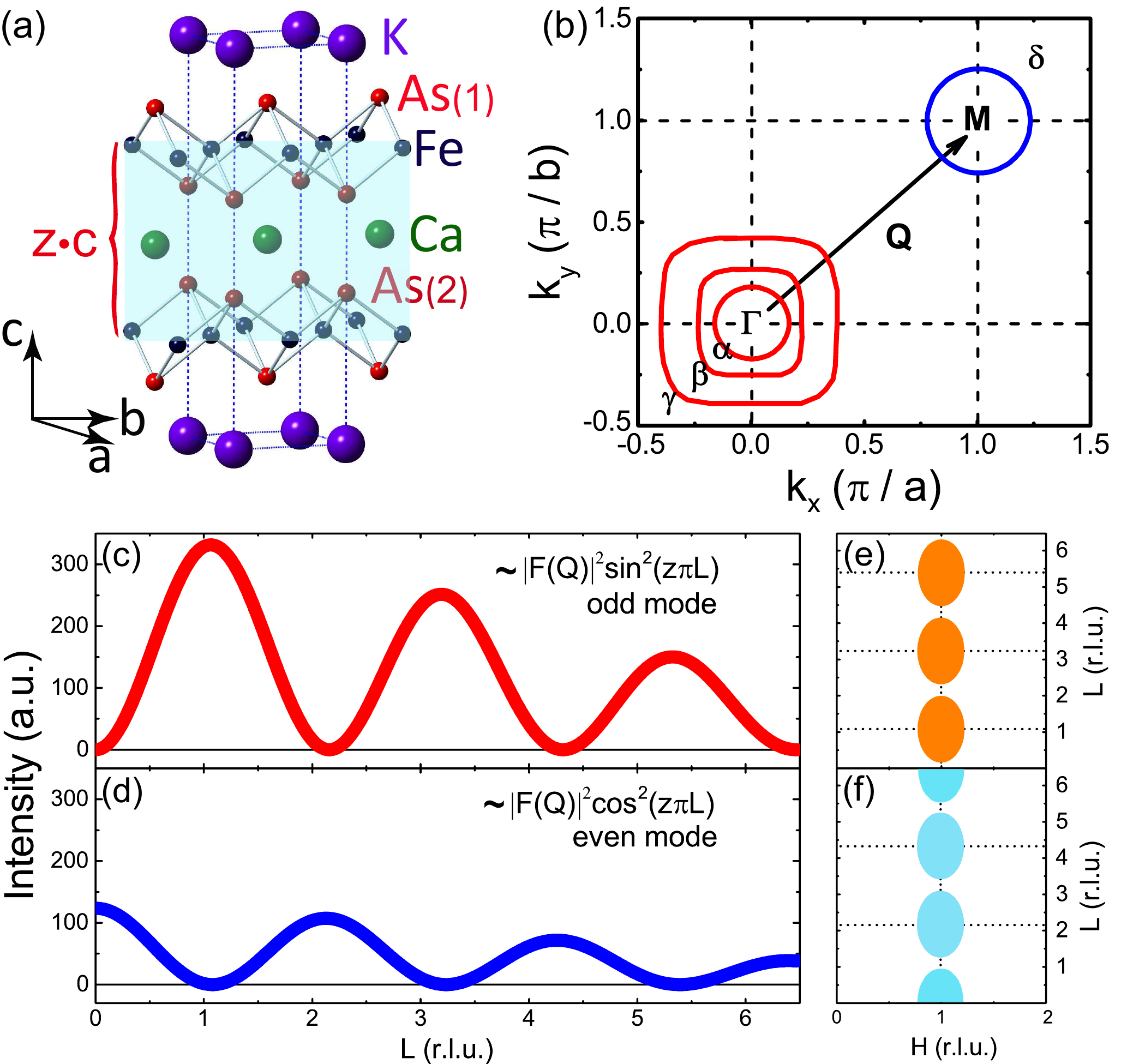}
\caption{
(a) Crystal structure of CaKFe$_4$As$_4$. (b) 2D Fermi surfaces with nesting wave vector ${\bf Q}$ from $\Gamma$ to $M$ point.
(c)$-$(f) Odd and even $L$ symmetries of the spin resonance.
}
\end{figure}

\begin{figure*}[t]
\includegraphics[width=0.9\textwidth]{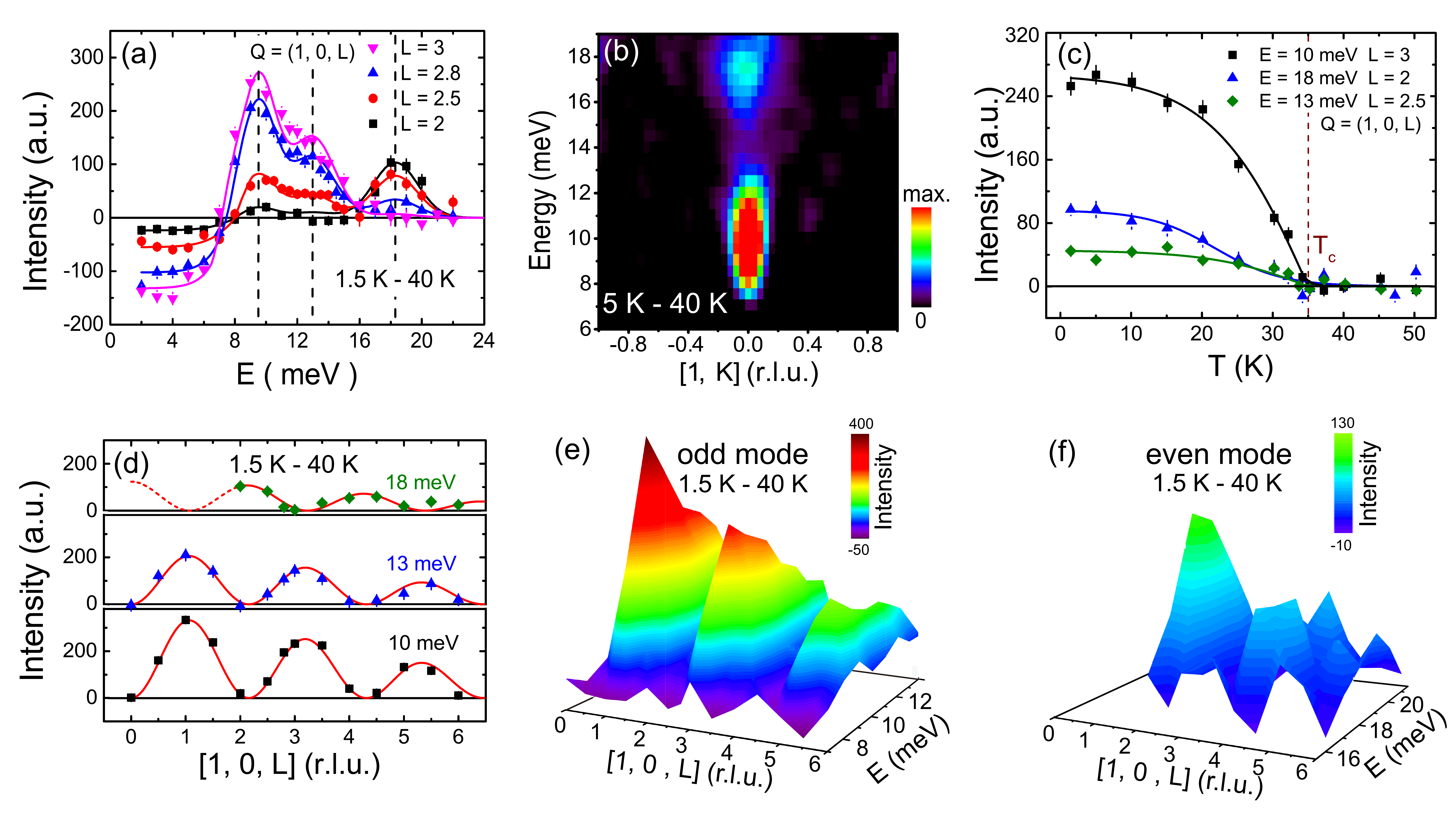}
\caption{(a) Energy dependence of the spin resonances at $Q = (1, 0, L)$. The solid lines are guides to eyes. (b) 2D slice in $E$ vs $K$ of the spin resonances. (c) Temperature dependence and (d) $L$ dependence of three resonance modes at $E=$ 10, 13, 18 meV.  The red solid and dashed lines are fitting results by $\mid F(Q)\mid^2\sin^2(z\pi L)$  [or $\mid F(Q)\mid^2\cos^2(z\pi L)$] function. (e), (f) $L$ modulation of the odd resonance modes and the even resonance mode at different energy ranges.
 }
 \end{figure*}

We prepared high-quality single crystals of CaKFe$_4$As$_4$ using the self-flux method according to the previous reports \cite{aiyo2016,wrmeier2016,wrmeier2017,supplementary}. X-ray diffraction, resistivity, and magnetization measurements suggest our crystals have a very homogenous quality with sharp superconducting transitions around 35 K, where the potential impurity phases from CaFe$_2$As$_2$ or KFe$_2$As$_2$ (122 compound) are completely absent. Neutron scattering experiments were carried out using thermal triple-axis spectrometer EIGER at SINQ, PSI, Switzerland, with a fixed final energy $E_f=$ 14.7 meV and about 2 g ($\sim$200 pieces) of coaligned crystals \cite{stuhr}. Time-of-flight (TOF) neutron scattering experiments were carried out at 4SEASONS spectrometer (BL-01) at J-PARC, Tokai, Japan, with incident energy $E_i= 42 $ and 23 meV, $k_i$ parallel to the $c$ axis, chopper frequency $f=250$ Hz, and a total sample mass of about 4.3 g ($\sim$400 pieces) \cite{siki1,Kajimoto2011,Inamura2013}. The scattering plane was $[H, 0, 0] \times [0, 0, L]$, defined using the magnetic unit cell with 2-Fe atoms similar to that of the 122 parent compounds: $a_M= b_M= 5.45$ \AA, $c=12.63$ \AA, in which the wave vector ${\bf Q}$ at ($q_x$, $q_y$, $q_z$) is $(H,K,L) = (q_xa_M/2\pi, q_yb_M/2\pi, q_zc/2\pi)$ reciprocal lattice units (r.l.u.).  The collinear ($C$-type) AFM order similar to CaFe$_2$As$_2$ (or the noncollinear spin vortex phase \cite{Meier2018}), which is expected to form magnetic Bragg peaks at $Q=(1, 0, L)$ [and $Q=(0, 1, L)$] ($L=\pm1, \pm3, \pm5, ...$), does not exist based on our neutron diffraction experiments (Supplemental Material \cite{supplementary}).  Even so, the spin excitations still emerge around $Q=(1, 0)$, coinciding with the Fermi surface nesting vector from $\Gamma$ to $M$ point [Fig. 1(b)], similar to many other iron pnictides [Fig. 1(e) and 1(f)]\cite{pdai2015,inosov2016}.

Figure 2 gives the key results of this paper. After subtracting the intensity of spin excitations in the normal state ($T=40$ K) from $E=2$ to 22 meV at $Q = (1, 0, L)$ with $L$ in the range $2-3$, we can identify three spin resonance peaks in the superconducting state ($T=1.5$ K) at $E_R=$ 9.5 $\pm$ 0.5, 13 $\pm$ 0.5, 18.3 $\pm$ 0.5 meV, respectively [Fig. 2(a)].  The intensity distribution of all three peaks separates into two groups, as clearly shown by the TOF neutron experiments [Fig. 2(b)]. Although the 9.5 and 13 meV modes overlap with each other, it seems all three modes are energy resolution limited and nearly nondispersive along both the $K$ and $L$ directions. The temperature dependence of all three modes confirms their coupling to superconductivity: the intensity gain decreases like a superconducting order parameter, which ceases at $T_c=35$ K [Fig. 2(c)]. A spin gapped feature with intensity loss below $T_c$ is also found below $E=7$ meV. More interestingly, all three modes show strong but different $L$ dependences with the maximums at $L =3$ for the former two modes and $L =2$ for the latter one [Fig. 2(a)]. Thus, we have further measured the spin excitations over a large range of $Q=(1, 0, L)$ with $L = 0-6$, where those below $L=2$ cannot be measured for $E\geq16$ meV due to the scattering restriction. The results are summarized in Figs .2(d)$-$2(f). Obviously, two opposite symmetries along $L$ can be identified for maximum around $L=$ odd or even, much similar to the cases in bilayer cuprates YBCO and Bi2212 \cite{meschrig2006,spaihes2003,spaihes2004,ysidis2007,lcapogna2007}. In the metallic YBCO, both odd and even modes of spin resonance are found corresponding to the acoustic and optical spin waves in the AFM insulating phase \cite{dreznik1996}. Although there is no evidence for any optical branch from antiphase spin excitations in the paramagnetic CaKFe$_4$As$_4$, by simply considering the symmetric and antisymmetric combinations from the contribution of magnetically decoupled Fe-As bilayers similar to metallic YBCO \cite{supplementary,hffong2000}, we can obtain both odd and even $L$ symmetries of the spin excitations. Here, the intensity of two spin resonances at $E_R=$ 9.5 and 13 meV follows the $L$ modulation $\mid F(Q)\mid^2\sin^2(z\pi L)$ (so-called odd mode), and the one at high energy ($E_R=$ 18.3 meV) can be described by $\mid F(Q)\mid^2\cos^2(z\pi L)$ (so-called even mode) instead, with respect to the distance ($zc$) between the adjacent Fe-Fe planes within the Fe-As bilayer unit [Fig. 1(a)] \cite{supplementary}. The data points agree very well with such sine-squared (cosine-squared) modulation, as shown in Fig. 2(d). Here, we have $z=0.4636$ for the unique bilayer structure due to the shift of the intermediate FeAs layer out of their high-symmetry positions ($z=$ 0.5) \cite{aiyo2016,wrmeier2016,wrmeier2017}. Consequently, the peak positions actually shift to nonintegral $L$ in comparison with the high-symmetric structure, such as $L$ = 1.08, 3.24, and 5.39 (odd mode) or $L$ = 2.16, 4.31, and 6.47 (even mode), etc [Fig. 1(c)$-$(f)]. Unlike the weak $L$ modulation of spin resonance in 122-type iron pnictides \cite{schi2009,zhang2011,chlee2013}, the minimum intensity at each valley here is near zero \cite{supplementary}.

Figure 3 summarizes the intensity distribution of the spin resonances and spin gap within $[H, K]$ plane. All three resonance modes and the spin gap follow Gaussian line shapes around $Q = (1, 0, L)$. While both the intensity loss at 3 meV and the intensity gain at 10 meV look like ellipses elongated along the $H$ direction, similar to the hole-doped Ba$_{1-x}$K$_x$Fe$_2$As$_2$ \cite{zhang2011}, the 13 and 18 meV resonant modes are more like circles in the $[H, K]$ plane. The peak width at half maximum of the intensity is determined by the dispersion of the paramagnetic excitation, and the relative intensity depends on the energy transfer coupled with the $L$ position in the TOF neutron scattering experiment when $k_i\parallel c$.

\begin{figure}[t]
\includegraphics[width=0.45\textwidth]{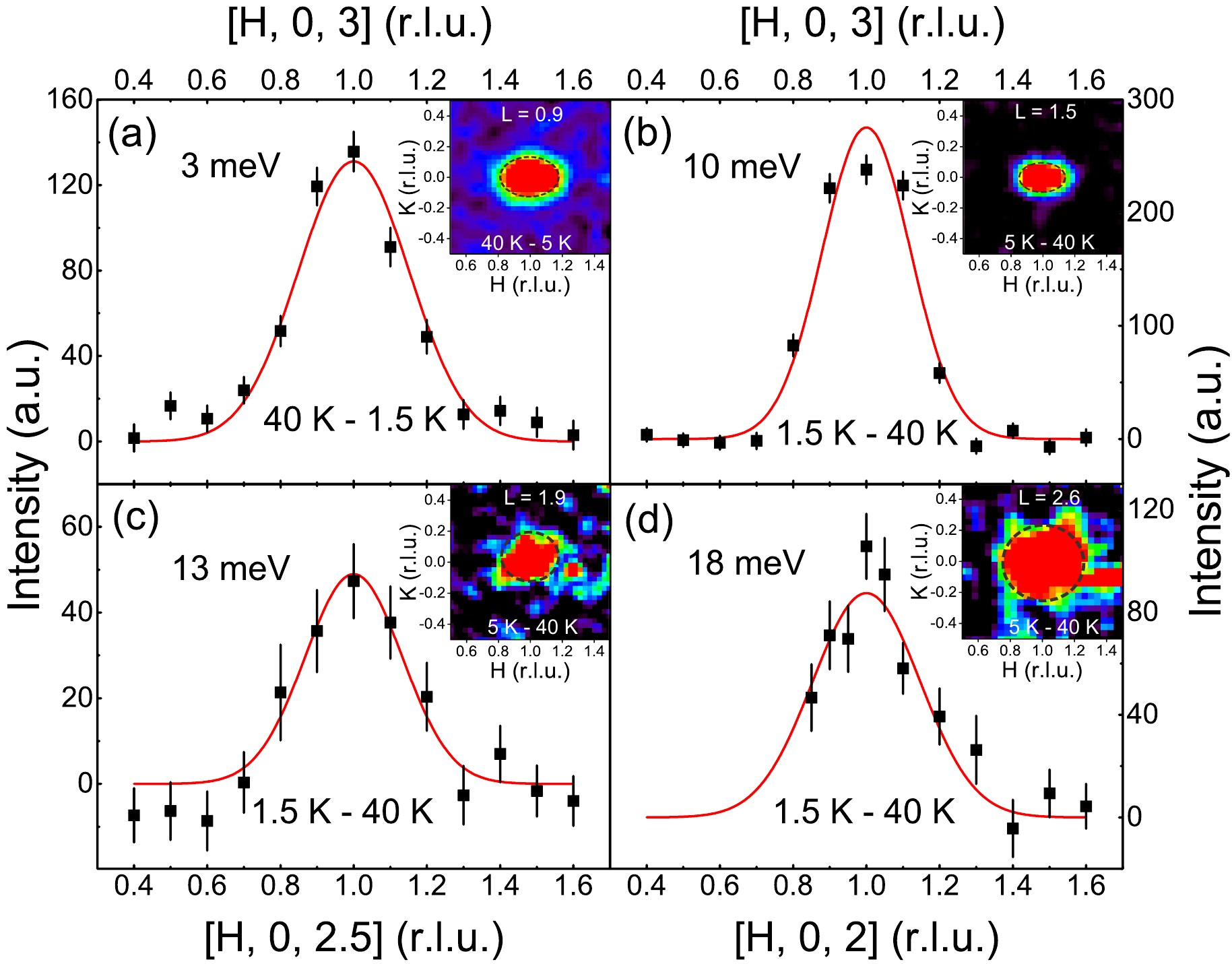}
\caption{ Constant-energy scans along the $H$ direction for (a) the spin gap at 3 meV and (b)$-$(d) three resonance modes at 10, 13, 18 meV with intensity differences below and above $T_c$. The red solid lines are Gaussian fits to the data. (Insets) 2D slices at half maximum of the intensity with identical energy transfer but different $L$s \cite{supplementary}.
 }
\end{figure}

The triple modes of spin resonance in CaKFe$_4$As$_4$ can be naturally explained by multiple pairing channels. Although the density functional theory calculations predict ten Fermi pockets (six hole bands and four electron bands) \cite{flochner2017}, the angle-resolved-photoemission-spectroscopy measurements reveal three hole pockets ($\alpha,\beta,\gamma$) around the $\Gamma$ point and one electron pocket ($\delta$) around the $M$ point, with large diversity of the superconducting gaps and 2D shapes of each pocket [Fig. 1(b)] \cite{dmou2016}. The observation of several full gaps and matched sizes of electron and hole pockets is consistent with the $s_\pm$ pairing scenario under Fermi surface nesting. Thus, three different values of the total superconducting gaps ($\Delta_{\rm tot}$) summed on the nesting hole and electron pockets should yield three modes of spin resonance at different energies \cite{supplementary}. It turns out that the resonance energy and the maximum intensity gain for each mode [$\Delta S(Q, \omega)$] show contrary linear dependence with $\Delta_{\rm tot}$ \cite{dmou2016,supplementary}, as shown in Fig.4 (a) and (b). In fact, a universal relationship $E_R/2\Delta=0.64$ was proposed among copper oxide, heavy fermion, and iron pnictide superconductors \cite{gyu2009}, where $2\Delta$ is twice the superconducting gap in the single band case. We then summarize the reported results about $E_R$ and $\Delta_{\rm tot}$ in Fig. 4(d) for iron-based superconductors \cite{yqiu2009,inosov2010,qureshi2012,zhang2013,wakimoto,qswang2016,pdjohnson,mywang2010,hqluo2013b,schi2009,dmou2016,zhang2011,chlee2013,nakayama2009,nakayama2011,chlee2016,kterashima2009,mwang2016,yzhang2012,jmaletz2014,hmiao2012,qqge2013,zswang2012,pustovit2016,jzhu2015,yzhang2011,xhniu2016}. The same linear scaling with $E_R=0.64\Delta_{\rm tot}$ can also describe these data together with our results of CaKFe$_4$As$_4$. Another well-known scaling behavior with $E_R=4.9k_BT_c$ may be still applicable in this new compound \cite{pdjohnson,txie2018}, only if considering the average resonance energy $E_R=12.5$ meV determined on a powder sample [Fig. 4(c)]\cite{kiida2017}. It should be noticed that all pockets are fully gapped in the superconducting state, and there is no evidence for gap modulation along $k_z$ or gap nodes in the spectroscopic investigations \cite{dmou2016,ryang2017,pkbiswas2016,kcho2017}. This agrees with the nondispersive feature of all three resonant modes and rules out the sign-changed gaps within a single Fermi pocket. Moreover, the orbital selective pairing could generate double resonant modes and possibly even $L$ modulation, as shown in the NaFe$_{1-x}$Co$_x$As system \cite{zhang2013b,zhang2013,wwang2017}. Unfortunately, further analysis on the orbital selection of pairing in CaKFe$_4$As$_4$ would be very difficult, given the equal occupation of Fe orbitals, including $d_{xz}$, $d_{yz}$, $d_{x^2-y^2}$, and $d_{z^2}$ \cite{flochner2017}.

\begin{figure}[t]
\includegraphics[width=0.46\textwidth]{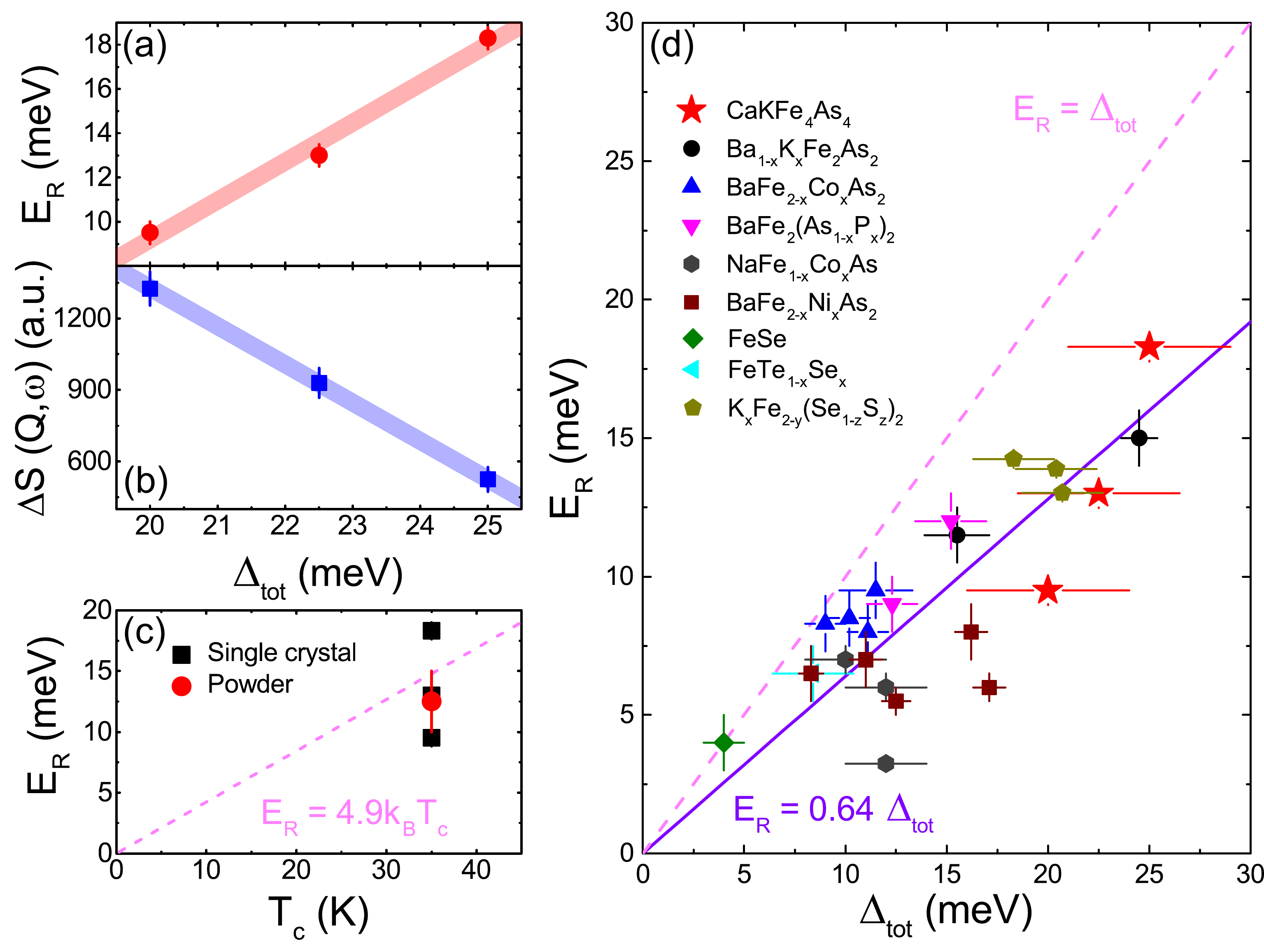}
\caption{
(a), (b) The linear relationship of $E_R$ and $\Delta S(Q, \omega)$ vs $\Delta_{\rm tot}$. (c) $E_R$ vs $T_c$ for CaKFe$_{4}$As$_{4}$ single crystal and powder samples under the linear scaling: $E_R$ = 4.9$k_B$$T_c$. (d) The linear scaling between $\Delta_{\rm tot}$ and $E_R$ for iron-based superconductors \cite{yqiu2009,inosov2010,qureshi2012,zhang2013,wakimoto,qswang2016,pdjohnson,mywang2010,hqluo2013b,schi2009,dmou2016,zhang2011,chlee2013,nakayama2009,nakayama2011,chlee2016,kterashima2009,mwang2016,yzhang2012,jmaletz2014,hmiao2012,qqge2013,zswang2012,pustovit2016,jzhu2015,yzhang2011,xhniu2016}. The dashed line marks $E_R=\Delta_{\rm tot}$, and the solid line is $E_R=0.64\Delta_{\rm tot}$ \cite{gyu2009}.
 }
\end{figure}

More importantly, the CaKFe$_4$As$_4$ compound actually is a bilayer system where the two Fe-As layers linked by Ca have a shorter distance along the $c$ axis than those linked by K for their different ionic radius \cite{aiyo2016,wrmeier2016,dmou2016,jcui2017,qpding2017} [Fig. 1(a)], thus the magnetic coupling within the bilayer unit is much stronger than the interbilayer interaction.  We also notice that the interlayer exchange coupling $SJ_c$ in CaFe$_2$As$_2$ is much larger (about 5.5 meV) \cite{jzhao2012} than that in BaFe$_2$As$_2$ (about 0.22 meV) \cite{jtpark2012}, and almost zero in KFe$_2$As$_2$ \cite{chlee2016b,mwang2013}, accompanied by the stretched Fe-As interlayer spacing with $0.5c=5.84$, 6.51, and 6.94 \AA\ \cite{nni2008,mrotter2008,hqluo2008,kkihou2010}, respectively. The distance of the Fe-Fe plane within one Fe-As bilayer of CaKFe$_4$As$_4$ is $0.4636c=5.855$ \AA\ , almost the same as the Fe-As interlayer spacing in CaFe$_2$As$_2$. Moreover, the energy difference between the odd (13 meV) and even (18.3 meV) spin resonance peaks is 5.3 meV, similar to $SJ_c$ in CaFe$_2$As$_2$. All these facts closely resemble those in the metallic YBCO with strong intrabilayer coupling $J_{\perp}\sim 10$ meV in the magnetically decoupled bilayers, where two spin resonance modes are found at 41 and 53 meV following the odd and even $L$ symmetries, respectively \cite{hffong2000,spaihes2004}. The existence of two spin resonance modes in YBCO and Bi2212 indicates that there is still an AFM coupling between Cu-O planes even in the superconducting state, which probably drives the bilayer systems to higher $T_c$ than the monolayer systems \cite{meschrig2006,ysidis2007}; whereas the multiband nature of CaKFe$_4$As$_4$ induces further splitting of the odd modes, thus generating triple peaks of spin resonance.

It should be noticed that the even mode of spin resonance in cuprates always has weaker intensity and higher energy than the odd mode. This is attributed to the presence of a threshold of the electron-hole spin flip continuum slightly below $2\Delta$, which supports the spin exciton scenario \cite{ysidis2007}. Although the dichotomy of theoretical descriptions of magnetism is still an unresolved issue in iron-based superconductors, the nearly isotropic spin resonance in most compounds basically agrees with the spin-1 exciton picture \cite{inosov2016,pdai2015,pdjohnson,txie2018}.  The multiple resonant modes remind us to recall the broadening and asymmetric spin resonance peak in many other iron-based superconducting systems, which are more likely induced by several overlapped odd and even modes due to small $SJ_c$ \cite{yqiu2009,inosov2010,qureshi2012,zhang2013,wakimoto,pdjohnson,mywang2010,hqluo2013b,schi2009,zhang2011,chlee2013,jzhao2013}.  If in analogy to the case of CaKFe$_4$As$_4$, the low-energy part of the resonance peak is probably filled with odd modes, while the even modes mostly contribute to the high-energy part. When changing $L$ from odd to even within one Brillouin zone, the overall peak position will naturally shift to higher energy \cite{supplementary}. This makes the resonant mode in appearance with a dispersion along the $L$ direction \cite{pdjohnson,mywang2010,hqluo2013b,schi2009,zhang2011,chlee2013,jzhao2013,mgkim2013,zhang2013b,zhang2013}. Finally, compared with our recent discovery of 2D spin resonance under three-dimensional Fermi surfaces in a 112-type iron-based superconductor \cite{txie2018},  it suggests that the resonance intensity is much more sensitive to the local magnetic couplings rather than the $k_z$ dependence of fermiology, even though the resonant energy is mostly coupled to superconducting gaps from itinerant electrons near the Fermi surfaces.

In summary, we have discovered strongly $L$-dependent triple spin resonance modes in the new iron-based superconductor CaKFe$_4$As$_4$ under multiple pairing channels. The resonance energies are below and proportional to the total superconducting gaps, consistent with the $s_\pm$ pairing mechanism. Both odd and even $L$ symmetries of the resonance intensity are found, which are attributed to the nondegenerate spin excitations from the Fe-As bilayer similar to the cuprate superconductors with the Cu-O bilayer. Our results suggest that the spin resonance in iron-based superconductors has an intrinsic common nature with cuprate superconductors, and the high-$T_c$ superconductivity in both families is strongly associated with local magnetic interactions coupled with itinerant electrons.

This work is supported by the National Key Research and Development Program of China (2017YFA0302903, 2017YFA0303103, 2016YFA0300502, 2015CB921300, 2017YFA0303100), the National Natural Science Foundation of China (11374011, 11374346, 11674406, 11334012, and 11674372), the Strategic Priority Research Program (B) of the Chinese Academy of Sciences (CAS) (XDB07020300), and the Key Research Program of the CAS (XDPB01). H. L. is grateful for the support from the Youth Innovation Promotion Association of CAS (2016004). This work is based on experiments performed at the Swiss Spallation Neutron Source (SINQ), Paul Scherrer Institute, Villigen, Switzerland. The neutron experiment at the Materials and Life Science Experimental Facility of J-PARC was performed under a user program (Proposal No. 2017A0051).

\clearpage
\appendix
\section{Supplementary Materials}
\begin{center}
{\bf A. SAMPLE CHARACTERIZATION}
\end{center}

We prepared high quality single crystals of CaKFe$_{4}$As$_{4}$ using self-flux method as previous reports \cite{Meier2016,Meier2017}. The sample photos and results of characterization are shown in Fig. S1 and Fig. S2. We co-aligned the crystals by a X-ray Laue camera (\emph{Photonic Sciences}) in backscattering mode with incident beam along $c$-axis on thin aluminium plates using \emph{CYTOP} hydrogen-free glue in $[H, 0, 0] \times [0, 0, L]$ scattering plane (Fig. S1). The crystalline quality was examined by single crystal X-ray diffraction (XRD) on a \emph{SmartLab} 9 kW high resolution diffraction system with Cu K$_{\alpha}$ radiation($\lambda$ = 1.5406 \angstrom) at room temperature ranged from $5\degree$ to $90\degree$ in reflection mode. Comparing to the CaFe$_2$As$_2$ system, the inequivalent position of Ca and K atoms in CaKFe$_{4}$As$_{4}$ leads to two different Fe-As distances below and above Fe-Fe plane and changes the space group from $I4/mmm$ to $P4/mmm$ with Fe-As bilayer structure (Fig.S1(e)) \cite{Meier2016,aiyo2016,dmou2016,flochner2017}, thus all odd and even Bragg peaks along $c$-direction are observed in XRD measurements due to the noncentrosymmetric structure. The sharp (0 0 $L$) peaks of X-ray diffraction in Fig. S1(b) indicate high $c$-axis orientation of our samples. The elastic neutron scattering results are shown in Fig. S1(c) and (d). There is no magnetic signal around $Q=(1, 0, 3)$ in this stoichiometric compound due to nearly zero intensity difference between 1.5 K and 70 K, suggesting no collinear (C-type) antiferromagnetism like CaFe$_{2}$As$_{2}$ or non-collinear spin vortex (hedgehog) structure like the Ni/Co doped case \cite{Meier2018}, where both of them have antiferromagnetic interlayer coupling along $c$-axes.

Figure S2(a) shows the normalized resistivity of our CaKFe$_{4}$As$_{4}$ crystals. The sharp superconducting transition width (less than 0.3 K), uniform $T_c$ and nearly identical normal state behaviors among 18 randomly selected samples indicate high quality and homogeneity of our samples. The normal state resistivity with high residual-resistivity-ratio RRR $\approx$ 15 also suggests high purity of our samples. Figure S2(b) shows the temperature dependence of the DC magnetic susceptibility for 4 typical crystals picked up from the resistivity measurements. The superconducting transitions of these samples are all very sharp with transition width less than 1 K. All these samples have full Meissner shielding volume ($4\pi\chi\approx -1$) at based temperature ($T = 2$ K). From the results of resistivity and DC magnetic susceptibility, we also can conclude that our CaKFe$_{4}$As$_{4}$ sample is purely homogeneous without any impurity phases from CaFe$_{2}$As$_{2}$ or KFe$_{2}$As$_{2}$, which may appear in the crystal growth process. For example, CaFe$_{2}$As$_{2}$ will induce a jump in the resistivity curve around 160 K for its structure transition, and KFe$_{2}$As$_{2}$ will result in another superconducting transition around 4 K in the magnetic susceptibility data \cite{Meier2016,Meier2017}.

\begin{center}
{\bf B. TRIPLE-AXIS NEUTRON SCATTERING}
\end{center}

Inelastic neutron scattering experiments were carried out using thermal triple-axis spectrometer EIGER at SINQ, PSI, Switzerland, with fixed final energy $E_f=$ 14.7 meV. The total sample mass is about 2 grams (about 200 pieces). Figure S3 shows energy scans at $Q = (1,0,L)$ ($L$ = 2, 2.5, 2.8, 3). After subtracting the intensity of spin excitations at the normal state ($T$ = 40 K) from $E =$ 2 to 22 meV at $Q = (1, 0, L)$ with $L = 2 \sim 3$, we can identify three spin resonance modes at the superconducting state ($T$ = 1.5 K) with center energies $E = 9.5 \pm 0.5, 13 \pm 0.5, 18.3 \pm 0.5$ meV, and a spin gap below 7 meV, respectively. The intensity gain for three resonant modes can be fitted by Gaussian peaks with modulated intensity and different width determined by energy resolution ($\delta E = 2.3, 2.8, 2.9$ meV). The maximums of the former two modes are at $L =$ 3 and the latter one is at $L =$ 2. It should be noticed that the strong peak above 16 meV in the raw data mainly comes from the phonon scattering of aluminum sample holders for both 1.5 K and 40 K. Since the phonon is almost unchanged for this temperature range (1.5 $\sim$ 40 K), we can identify the resonance mode around 18 meV by comparing the intensity below and above $T_c$.

Figure S4 summarizes constant-energy scans along $H$ direction for the gap energy 3 meV and the resonant modes at $E=$ 10, 13, 18 meV. There are some spurious scattering signal in the raw data from aluminium phonon and high order neutron scattering. After subtracting the 1.5 K (40 K for 3 meV) data by the 40 K (1.5 K for 3 meV) data, we have well-defined Gaussian peaks for all measured energies.

In order to figure out the $L-$dependence of the spin excitations, we have measured the spin excitations for the energies over a wide range of $\textbf{Q} = [1, 0, L]$ with $L = 0 \sim 6$. The $L$ dependence of the results for the gapped and resonant energies at $E=$ 3, 10, 13, 18 meV were presented in Fig. S5. Periodic modulations can be identified in the raw data of these $L$ scans, which is more clear in the the subtracted data between 1.5 K and 40 K. The modulation of the resonance around 18 meV has an opposite behavior with 10 and 13 meV intensities. This is also consistent with the results in Fig. S3(b, d, f, h). Figure S6 summarizes the intensity differences between 1.5 K and 40 K (resonance intensity) for energies from 7 to 21 meV, which can be clearly separated to two different groups fitted by harmonic functions. We define the \textquotedblleft odd mode\textquotedblright described by $\mid F(Q)\mid^2\sin^2(z\pi L)$, and \textquotedblleft even mode\textquotedblright described by $\mid F(Q)\mid^2\cos^2(z\pi L)$, respectively, where $F(Q)$ is the magnetic form factor of $\mathrm{Fe}^{2+}$, and $zc=5.855$ \AA\ ($z=$0.4636, $c=12.63$ \AA) is the distance between adjacent Fe-Fe planes within the crystalline block of Fe-As bilayer (Fig.S1(e))\cite{Pailhes1,Pailhes2}. It should be mentioned that $z=$0.4636 is not a free fitting parameter in our case, but obtained from the structure refinement of the CaKFe$_{4}$As$_{4}$ samples \cite{aiyo2016,dmou2016}. Because the translation symmetry along $c$-axes is broken in this compound, the intermediate FeAs-layers shift out of their high-symmetry positions, forming the bilayer structure with $z<$0.5 and two different Fe-As distances below and above the Fe-Fe plane (Fig.S1). The non-integral $L-$positions of maximum intensity of spin resonance do not mean it is incommensurate along $L$ direction, since the intensity modulations spread in the entire Brillouin zone (Fig.S5,Fig.S6).

\begin{center}
{\bf C. TIME-OF-FLIGHT NEUTRON SCATTERING}
\end{center}

 Time-of-flight (TOF) neutron scattering experiments were carried out at 4SEASONS spectrometer (BL-01) at J-PARC, Tokai, Japan\cite{Kajimoto2011,Inamura2013}. The incident energies $E_i$ = 42 and 23 meV with $k_i$ in parallel to $c$-axis, chopper frequency $f=$ 250 Hz. Thus the energy transfer $E$ is mostly in coupled with $L$ for the quasi-2D lattice structure. The total mass of co-aligned samples is about 4.3 grams (about 400 pieces, see Fig. S1(a)). Figure S7 gives the 2D slices of $E$ vs. $K$ for $T=$ 5 K and 40 K measured by TOF experiments with $E_i$ = 42 meV, where both the $\mid Q \mid-$dependent background from phonon scattering and constant background from incoherent scattering are subtracted. The spin excitations significantly increase below $T_c$ around $E=$ 10 and 18 meV. By directly subtracting the raw data at 40 K in the normal state from the 5 K data, we get the 2D slice in $E$ vs. $K$ of the spin resonances at 5 K (superconducting state)(Fig. 2(b)).

In our TOF experiments, the scattering plane is $[H, 0, 0] \times [0, 0, L]$ under $k_i \parallel c$, thus energy transfer $E$ is mostly in coupled with $L$ for the quasi-2D lattice structure. To show the 2D constant-energy slices in [$H, K$] plane for the spin gap at 3 meV and three modes of spin resonance at 10, 13, 18 meV (Fig.S8), we integrated the signal in a narrow energy window $E \pm \Delta E$ as shown in each panel, which is corresponding to a specific $L$ indicated by the inserts of Fig.3. The two bright specks in Fig.S8(a) are spurious from the contamination of (1, 1, 1) Bragg peaks accidentally hitting on the detector, where is actually no signal due to the full spin gap. From these 2D color maps we can get the clear information of lineshape of the spin gap and the spin resonances.  Four ellipses elongated along radial direction are found for the paramagnetic excitations. The four-fold slices are shown in Fig.3 with normalized intensity cutoff at the half maximum. The elongated ellipses along $H$ direction at $E=$10 meV agree with the mismatch between $\gamma$ band and $\delta$ band.

\begin{center}
{\bf D. SUPERCONDUCTING GAP AND SPIN RESONANCE ENERGY}
\end{center}

In the single-band system such as cuprates, the spin resonance is believed to be a collective mode of spin-1 singlet-triplet excitations in the superconducting state, thus the spin resonance energy $E_R$ is below the pair-breaking energy $2\Delta$ (twice the superconducting gap) due to creation of particle-hole pairs \cite{gyu2009}. However, the iron-based superconductors are multi-band systems, different gaps (probably with different sign) are observed on different pockets of Fermi surface. In the itinerant picture, the sign-reversed $s-$wave ($s\pm$) superconductivity establishes via the Fermi surface nesting between the hole pocket and electron pocket (Fig.S9(a)) \cite{korshunov2008,avchubukov2008,mazin2009}. Here, the pair-breaking energy is the sum of superconducting gaps on the nesting pockets $\Delta_{tot}=|\Delta_k|+|\Delta_{k+Q}|$, or $\Delta_{tot}=|\Delta_h|+|\Delta_p|$, where $Q$ is the nesting vector, $\Delta_k$ or $\Delta_h$ is the gap on hole sheet, $\Delta_{k+Q}$ or $\Delta_p$ is the gap on electron sheet. The spin resonance mode is an in-gap bound state determined by the coherence factor $[1-\Delta_k\Delta_{k+Q}/E_kE_{k+Q}]/2$, where $E_k$ ($E_{k+Q}$) is the quasiparticle energy \cite{tamaier2009,qswang2016}. At the Fermi level, $E_k=\sqrt{\varepsilon^2_k+\Delta^2_k}=|\Delta_k|$, the gap sign should be reversed ($\Delta_k\Delta_{k+Q}/|\Delta_k||\Delta_{k+Q}|=-1$) to produce a finite intensity of the resonance with $E_R < \Delta_{tot}$. Alternatively, a non-resonance broad peak above $2\Delta$ (here $\Delta_k=\Delta_{k+Q}=\Delta$) may also emerge under the conventional sign-preserved ($s_{++}$) pairing picture \cite{sonari2010,sonari2012}, due to the redistribution of the magnetic spectral weight when cooling down to the superconducting state. In some iron chalcogenide superconductors (e.g. K$_x$Fe$_{2-y}$Se$_2$, (Li$_{1-x}$Fe$_x$)OHFeSe, mono-layered FeSe thin film), there are only electron pockets \cite{qfan2015,dfliu2012,slhe2013,sytan2013,jfhe2014,lzhao2016,xliu2014}. The spin resonance mode is then observed at the nesting wave vector $Q$ linked by two electron pockets \cite{jtpark2011,gfriemel2012}, and a sign change of the gap between the inner and outer electron pocket is proposed from the results of quasi-particle interference measurements (Fig.S9(b)) \cite{zydu2016,zydu2018}. In this case, the pair-breaking energy is still valid to represent by the total superconducting gaps $\Delta_{tot}=|\Delta_k|+|\Delta_{k+Q}|$ summed on the two nesting electron pockets. In both cases with $s\pm$ pairing, the spin excitations form a sharp resonant peak at energy $E_R$ below $\Delta_{tot}$.

For our case in CaKFe$_4$As$_4$, $\Delta_{tot}$ is the total superconducting gaps summed on the nesting pair of hole band ($\alpha,\beta,\gamma$) and electron band ($\delta$), which is determined by Angle-Resolved-Photoemission-Spectroscopy (ARPES) experiments \cite{dmou2016}. We list them below and compare with the resonance energies in Tabel.S1.  The ratio $E_R/\Delta_{tot}$ of three resonant modes is around 0.6, which is consistent with the $s\pm$ pairing mechanism. We also notice that the best nesting condition ($\beta$ to $\delta$) results in the largest superconducting gaps and the highest resonance energy, and the elongated ellipses in the 2D slices of spin resonance in $[H, K]$ plane at $E=10$ meV agree with the mismatch between $\gamma$ band and $\delta$ band (Fig.3). Besides the conventional $s\pm$ state with sign change between hole and electron pockets, a recent theory also predicts a C-state pairing symmetry with an additional sign change within hole or electron pockets in CaKFe$_4$As$_4$ \cite{flochner2017}. When the Coulomb repulsion $U$ is weak, some interband interactions may change sign and become weakly attractive. Thus the C-state could induce a weak enhancement of the spin fluctuation below $T_c$ around relatively high energy above $\Delta_{tot}$ and a near-nodal behavior of the quasiparticle excitations at some electron pockets. However, it turns not exactly the case in our neutron experiments, since the high energy even mode of spin resonance has resolution-limited peak width and $E_R < \Delta_{tot}$ and no gap nodes are observed in the spectroscopic investigations \cite{dmou2016,ryang2017,pkbiswas2016,kcho2017}.

\begin{table}
Table S1. Superconducitng gaps and resonance energies in CaKFe$_4$As$_4$.\\
\begin{center}
\begin{tabular}{|c|c|c|c|c|c|}
 \hline \hline
hole   & $\Delta_h$   &  $\Delta_p(\delta)$      & $\Delta_{tot}$  &  $E_R$ &  $E_R/\Delta_{tot}$\\
pocket   & (meV)   &  (meV)       &  (meV) &  (meV) &    \\
  \hline
$\alpha$     & 10.5        &       & 22.5 & 13  & 0.58  \\
$\beta$      & 13       & 12     & 25   & 18.3   & 0.73  \\
$\gamma$    & 8       &       & 20      & 9.5  & 0.48 \\
\hline \hline
\end{tabular}
\end{center}\label{tab:tableI}
\end{table}

\begin{center}
{\bf E. $L-$SYMMETRY OF THE SPIN RESONANCE}
\end{center}

In the insulating bilayer cuprate YBa$_2$Cu$_3$O$_{6}$ (YBCO), the acoustic magnons are defined as the low energy sector of the spin excitation spectrum which evolves out of in-phase precession modes of spins in directly adjacent layers, while the optical magnons are the higher energy sector evolving out of antiphase spin excitations \cite{dreznik1996}. The former have odd symmetry and the latter have even symmetry under exchange of the two layers. In the metallic phase of YBa$_2$Cu$_3$O$_{6+\delta}$, such acoustic and optical spin waves will develop into odd and even spin excitations with two corresponding spin resonance modes below $T_c$, so called odd and even modes of spin resonance.

Except for some magnetically ordered iron chalcogenides, so far there is no evidence for optical magnons in parent compounds of iron-based superconductors \cite{pdai2015}. However, regardless the acoustic- or optical-like magnons, we can still deduce the odd and even symmetries from the non-degenerate inter-layer magnetic excitations within decoupled bilayer similar to YBCO \cite{hffong2000}. We assume the eigenstate of the Fe-As layer $n$ as $\mid{n}\rangle$ ($n=1, 2$):
\begin{equation}
\left\{
\begin{array}{ccc}
\hat{z}\mid{1}\rangle =+d/2\mid{1}\rangle,\\
\\[1pt]
\hat{z}\mid{2}\rangle =-d/2\mid{2}\rangle,\\
\end{array}
\right.
\end{equation}
 where $d=zc$ is the distance between two adjacent Fe-Fe layers. We can then define symmetric and antisymmetric combinations of the states centered on the two layers:
\begin{equation}
\left\{
\begin{array}{ccc}
\mid{s}\rangle =(\mid{1}\rangle+\mid{2}\rangle)/\sqrt{2},\\
\\[1pt]
\mid{a}\rangle =(\mid{1}\rangle-\mid{2}\rangle)/\sqrt{2}.\\
\end{array}
\right.
\end{equation}
With momentum transfer $Q$ along $c-$axis, the spin excitations characterized by the transitions between $\mid{s}\rangle$ and $\mid{a}\rangle$ states are given by the following superposed states:
 \begin{equation}
\left\{
\begin{array}{ccc}
\langle s\mid e^{iQ\hat{z}}\mid{s}\rangle = \langle a\mid e^{iQ\hat{z}}\mid{a}\rangle=\cos(Qd/2)\ (\mathrm{even}),\\
\\[1pt]
\langle s\mid e^{iQ\hat{z}}\mid{a}\rangle = \langle a\mid e^{iQ\hat{z}}\mid{s}\rangle=i\sin(Qd/2)\ (\mathrm{odd}).\\
\end{array}
\right.
\end{equation}
Here $Qd/2=2\pi L/c \times zc/2=z\pi L$ . Thus the dynamic spin susceptibility of odd and even excitations can be described by:
\begin{equation}
\left\{
\begin{array}{ccc}
\chi''_{\mathrm{odd}}(Q,\omega)\sim \sin^2(z\pi L),\\
\\[1pt]
\chi''_{\mathrm{even}}(Q,\omega)\sim \cos^2(z\pi L).\\
\end{array}
\right.
\end{equation}
The spin-spin correlation function $S(Q,{\omega})$ is related to the local susceptibility: $S(Q,{\omega})=(1+n(\omega))\chi''(Q,\omega)/\pi g^2\mu_B^{2}$, where $1+n(\omega)$ is the Bose factor. The cross section $d^2\sigma/(d\Omega dE)$ should be further normalized by the square of Fe$^{2+}$ magnetic form factor $\mid F(Q)\mid^2$ and Debye-Waller exponent $\exp(-2W(Q))$.

\begin{center}
{\bf F. COMPARISON WITH OTHER PNICTIDES}
\end{center}

 The presence of both even and odd modes of spin resonance are seen specifically in CaKFe$_4$As$_4$ for the following reasons. Firstly, this compound has a stoichiometric superconductivity without any magnetic order or any disorder from the dopants which may induce complexity in the pairing process. Secondly, the Fermi surfaces are nearly 2D with a range of diameters, resulting in multiple nesting conditions. Thirdly, the superconducting gaps are divergent from each Fermi sheet, therefore different modes of the spin resonance from the multiple nesting can be separately identified at different energies. Finally, the inequivalent position of Ca and K atoms leads to a bilayer symmetry similar to YBCO. The split between odd and even resonant modes (about 5.3 meV) are determined by the strong intra-bilayer coupling ($SJ_c=$ 5.5 meV in CaFe$_2$As$_2$).

In other iron-based superconducting systems, the spin resonance peak is thus likely including several overlapped odd and even modes due to small $SJ_c$ \cite{pdai2015}. Since the maximum intensity of spin resonant mode ($\Delta S$) decreases with increasing resonance energy ($E_R$) (Fig.4(b)), if neighboring odd and even modes are overlapped with each other due to small interlayer coupling, an asymmetric broad peak with a long tail at high energy part then is expected, instead of several individual peaks. In fact, this feature has been already observed in many iron-based superconductors \cite{pdai2015,inosov2016,pdjohnson}. Here we take three systems for example: BaFe$_{1.925}$Ni$_{0.075}$As$_2$, Ba(Fe$_{1-x}$Ru$_{x}$)$_2$As$_2$ and BaFe$_2$(As$_{1-x}$P$_x$)$_2$ (Fig.S10, Fig.S11 and Fig.S12) \cite{mywang2010,jzhao2013,chlee2013}. In the underdoped BaFe$_{1.925}$Ni$_{0.075}$As$_2$ with stripe-type magnetic order, a clear odd $L-$modulation of the spin excitations is observed. The spin resonance peak is indeed asymmetric with a long tail at high energy part, and the peak center shifts to lower energy when increasing $L$ from 0 to 1 \cite{mywang2010}. Similar behaviors exist in the Ba(Fe$_{1-x}$Ru$_{x}$)$_2$As$_2$ system, no matter in the magnetically ordered state (underdoped regime) or paramagnetic state (optimally doped level) \cite{jzhao2013}. If we suppose the even modes are hidden in the high energy part above the peak center with weak intensity, when $L$ increases from 0 to 1, the intensity at high energy part (even modulation) decreases, while the intensity at low energy part (odd modulation) increases. Then it looks like the resonance peak center shifts to lower energy, resulting in a dispersion of the resonance mode. By digging out the data of BaFe$_2$(As$_{1-x}$P$_x$)$_2$ in Fig.S12 (a), we indeed find opposite $L$ dependence of resonance intensity at 9 meV and 12 meV. To clarify this issue, further neutron scattering experiments studied on the spin resonance over a large range of $L$ need to be done, the polarized neutron analysis may also help to identify the overlapped modes with different orbital character.

\clearpage
\begin{figure*}[t]
\renewcommand\thefigure{S1}
\includegraphics[width=0.8\textwidth]{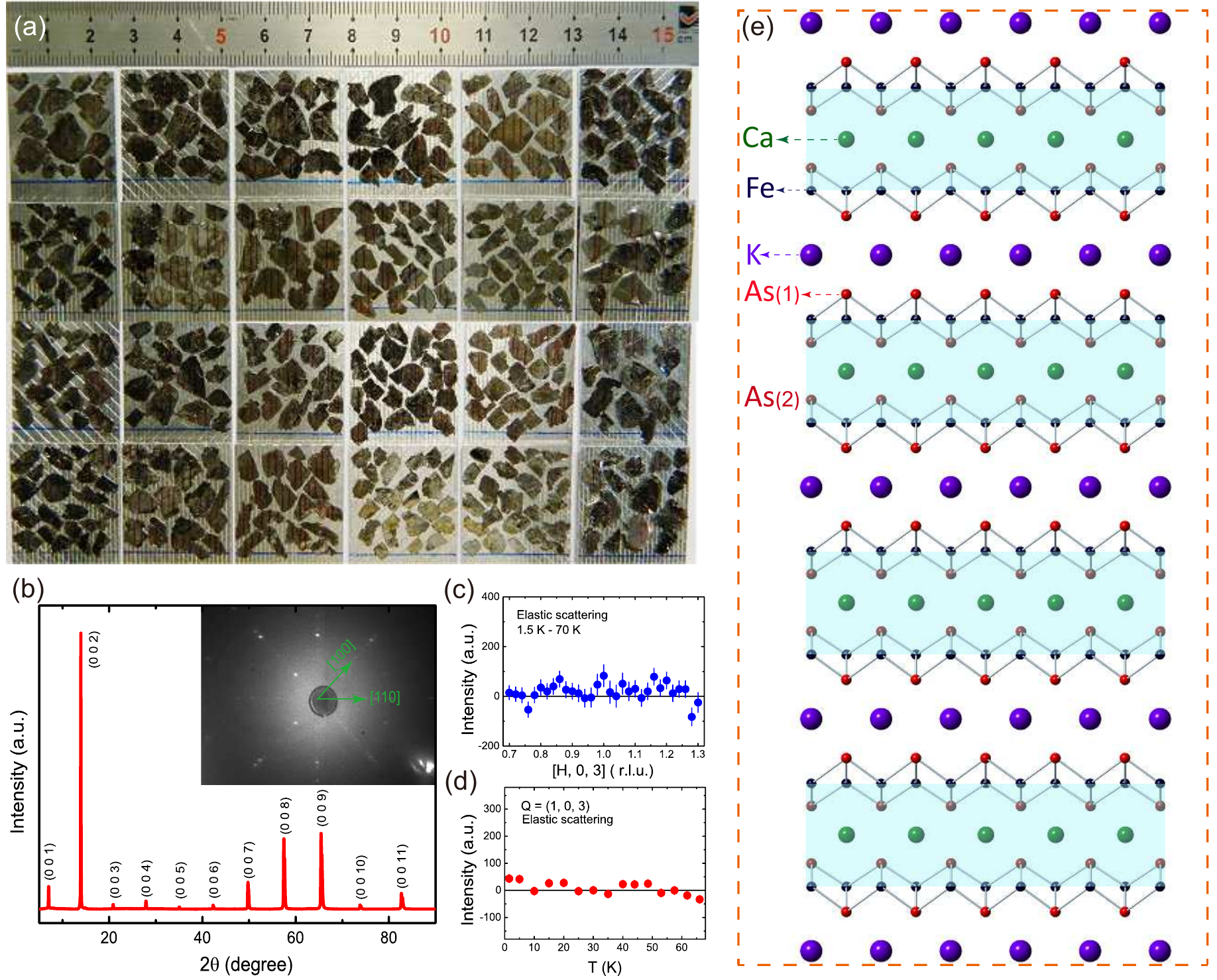}
\caption{
(a) Photos of the co-aligned CaKFe$_4$As$_4$ crystals for neutron scattering experiments. (b) X-ray diffraction pattern of CaKFe$_4$As$_4$ single crystal at room temperature, the inset is a Laue photo of CaKFe$_4$As$_4$ single crystal, the high symmetry directions [100] and [110] are indicated by green arrows. (c, d) $Q$ and temperature dependence of the elastic neutron scattering at $Q = (1,0,3)$ . (e) The crystalline blocks with FeAs bilayers.
}
\end{figure*}

\begin{figure*}[t]
\renewcommand\thefigure{S2}
\includegraphics[width=0.85\textwidth]{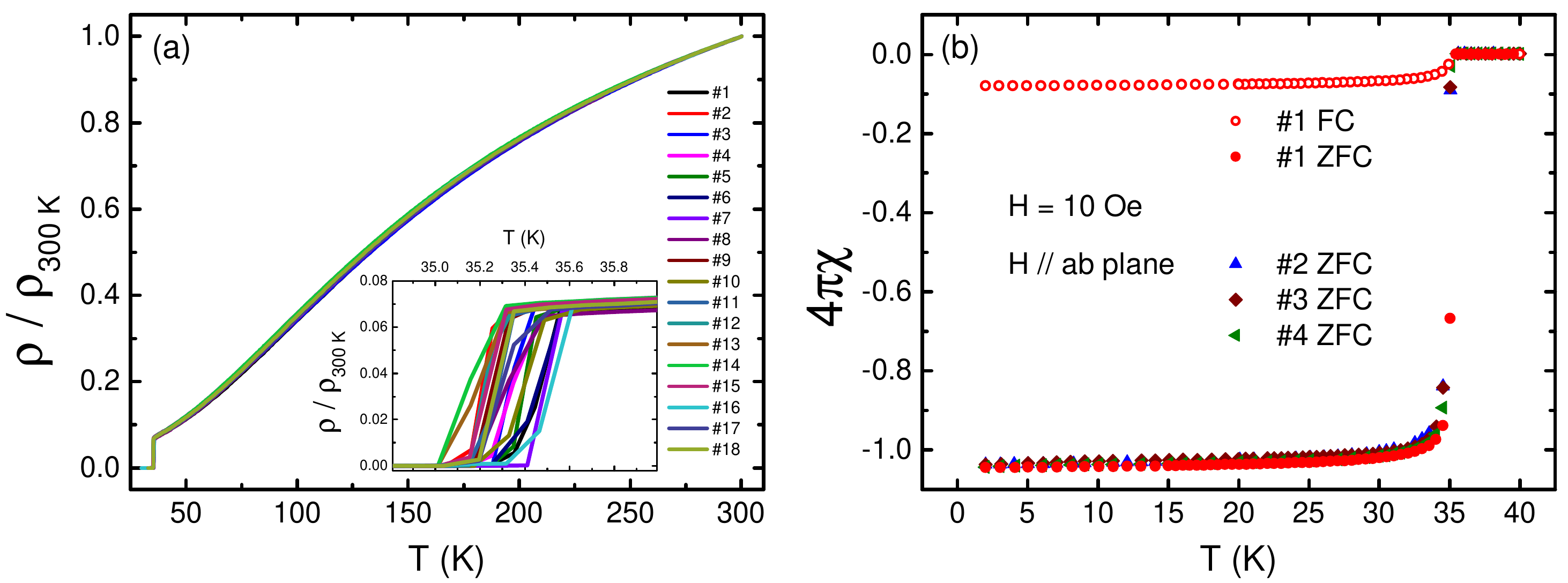}
\caption{Superconductivity of CaKFe$_4$As$_4$ single crystals: (a) Temperature dependence of the resistivity, all the data is normalized by the resistivity at 300 K; (b) Temperature dependence of DC magnetic susceptibility.
 }
 \end{figure*}

\begin{figure*}[t]
\renewcommand\thefigure{S3}
\includegraphics[width=0.8\textwidth]{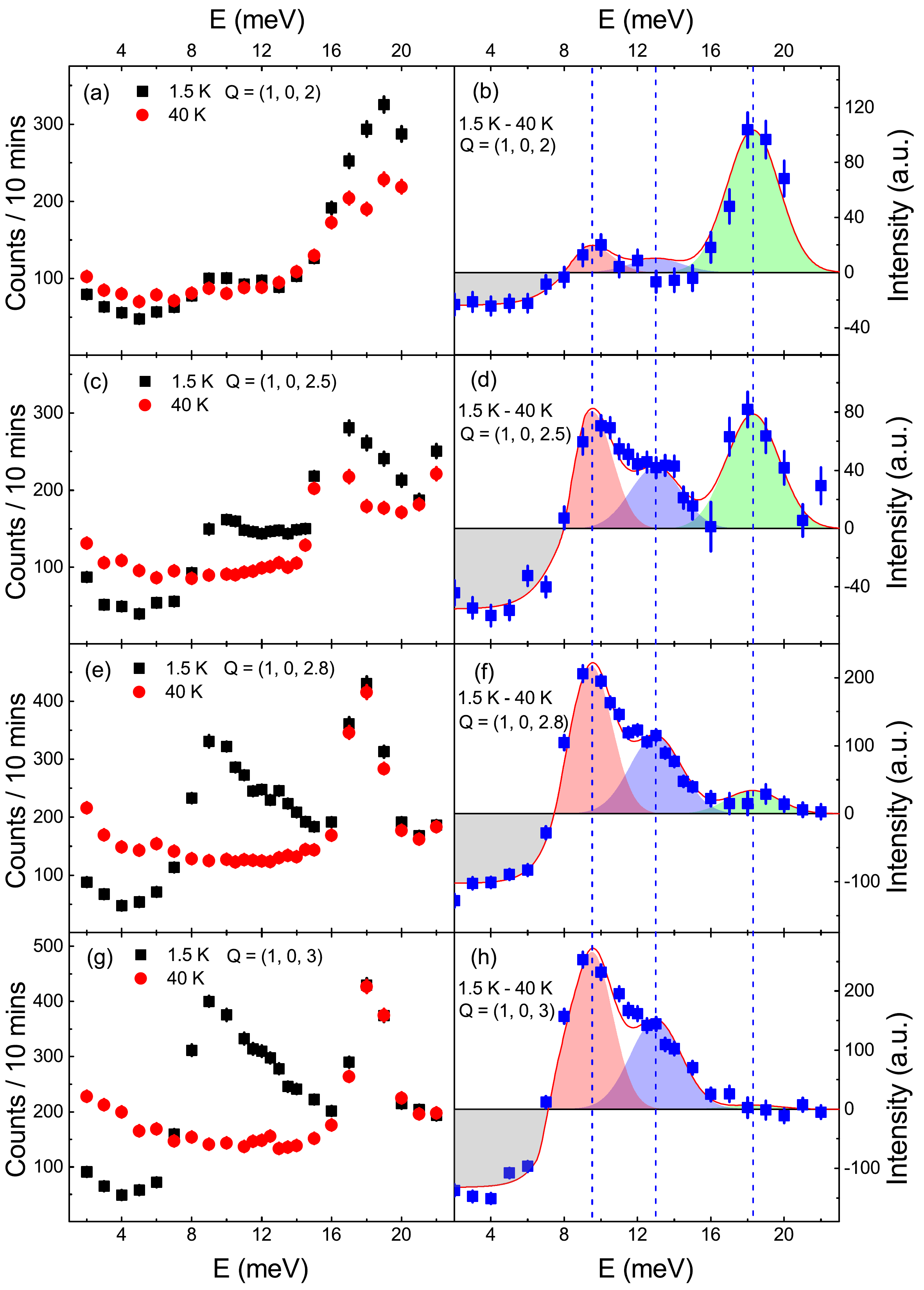}
\caption{ Energy scans for scattering below and above $T_c$, and their differences at $Q=$ (1, 0, 2), (1, 0, 2.5), (1, 0, 2.8), (1, 0, 3). The solid lines are guides to the eyes. The shadow area are gaussian fits for the intensity gain of each resonant modes and the intensity loss of the spin gap. The dashed lines indicate three resonant modes around $E=$ 9.5 meV, 13 meV, 18.3 meV, respectively.
 }
\end{figure*}

\begin{figure*}[t]
\renewcommand\thefigure{S4}
\includegraphics[width=0.8\textwidth]{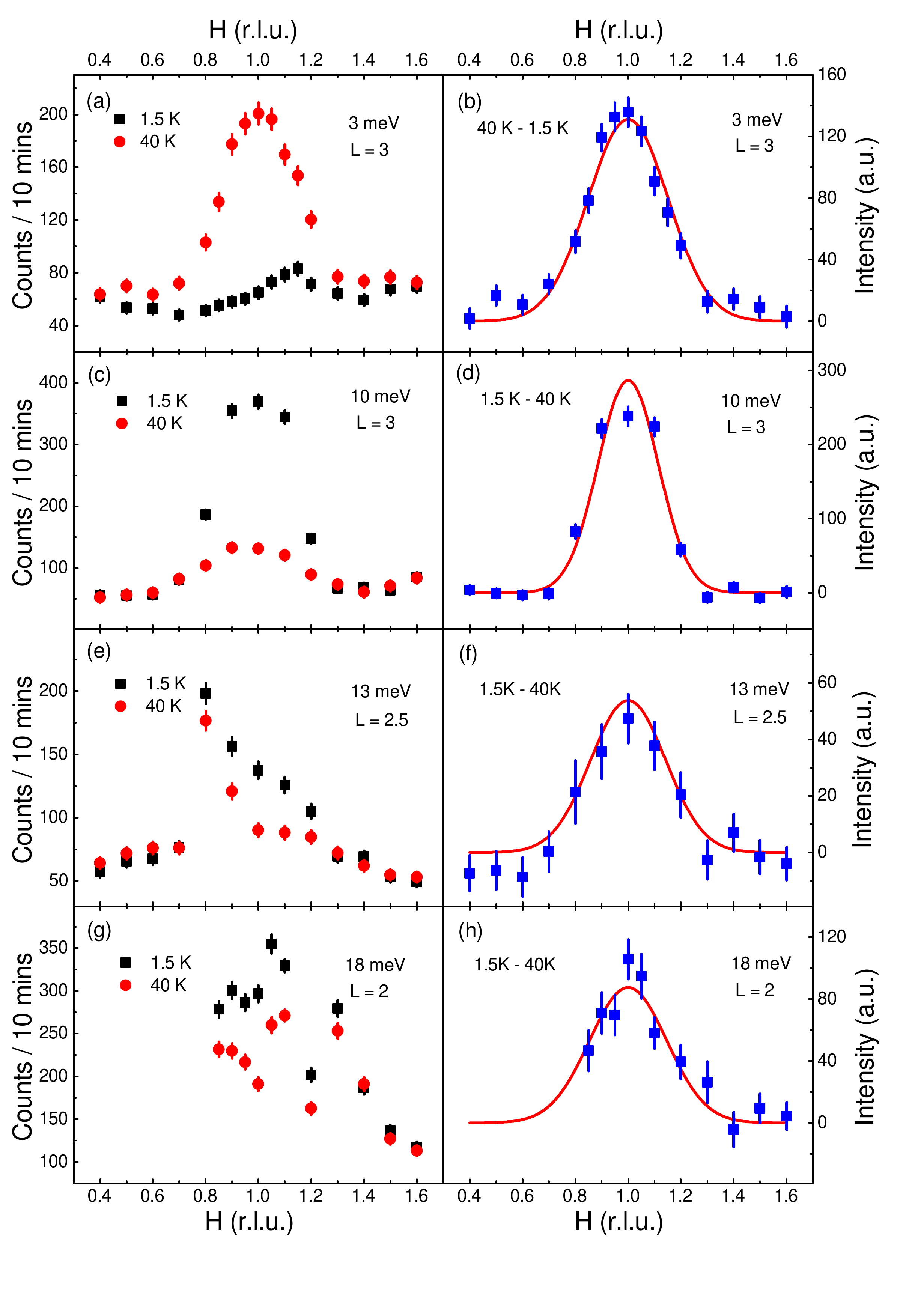}
\caption{
Constant-energy scans along $H$ direction for the spin gap at $E = 3$ meV and three resonant modes at $E = 10, 13, 18$ meV with $L = 2 \sim 3$. The red solid lines are Gaussian fits.
 }
\end{figure*}

\begin{figure*}[t]
\renewcommand\thefigure{S5}
\includegraphics[width=0.8\textwidth]{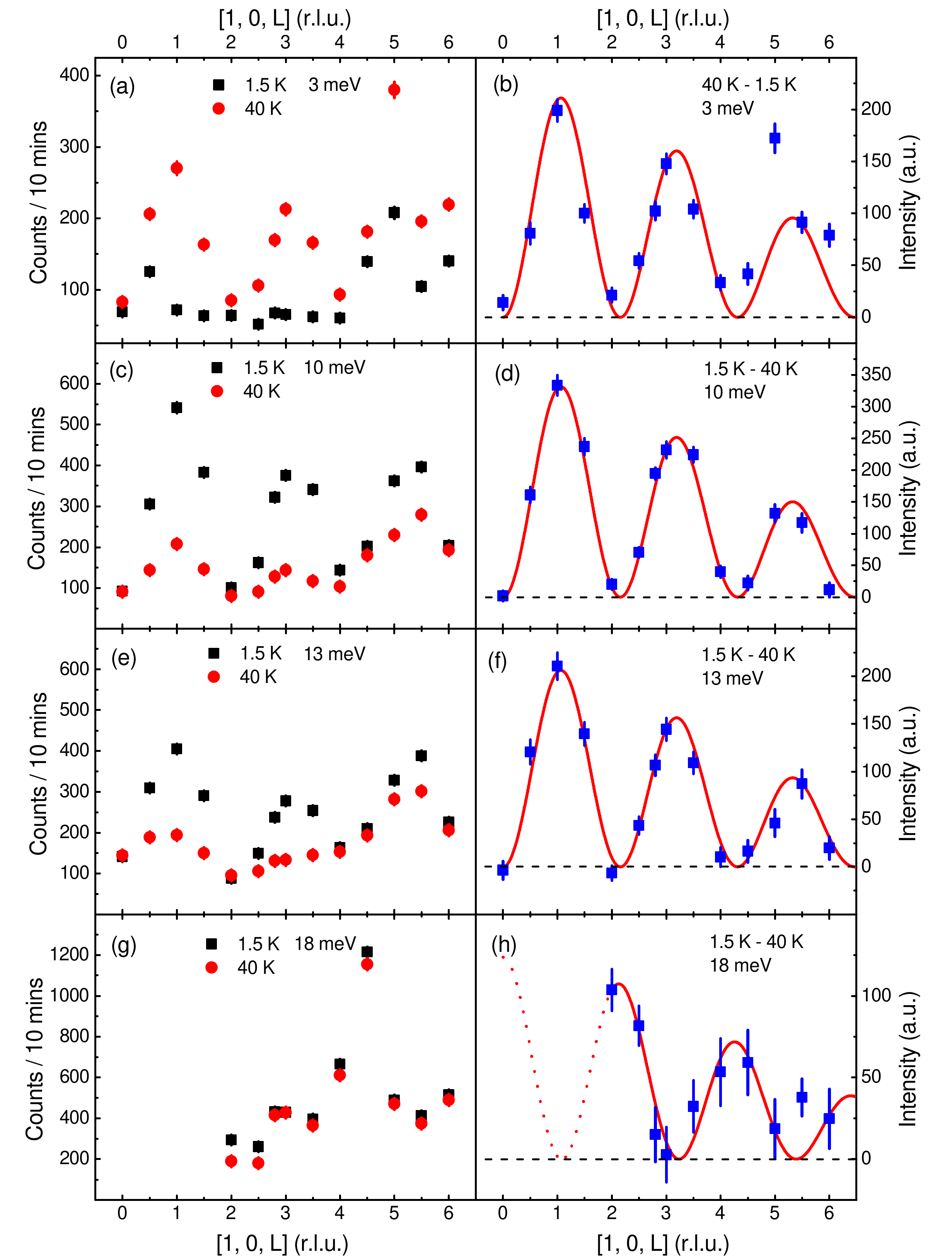}
\caption{ Constant-energy scans along $L$ direction for the spin gap at $E = 3$ meV and three resonant modes at $E = 10, 13, 18$ meV. The solid lines in (b, d, f) are fitting results by $\mid F(Q)\mid^2\sin^2(z\pi L)$ (odd mode), while the red line in (h) is fitting result by $\mid F(Q)\mid^2\cos^2(z\pi L)$ (even mode).
 }
\end{figure*}

\begin{figure*}[t]
\renewcommand\thefigure{S6}
\includegraphics[width=0.8\textwidth]{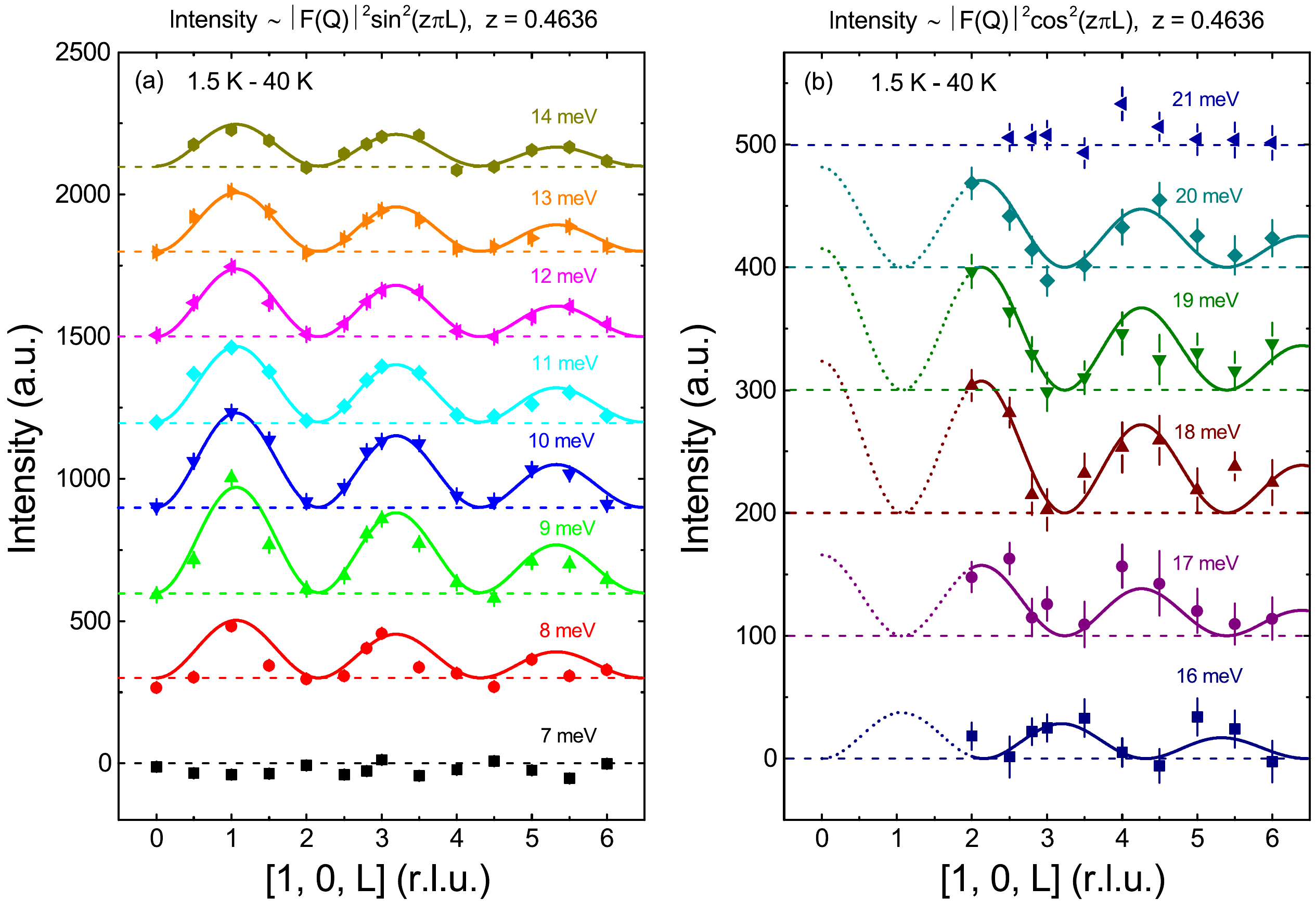}
\caption{
(a) and (b)Intensity differences of the scans along $L$ direction between 1.5 K and 40 K with $E = 7 \sim 21 $ meV . The solid lines are fitting results by $\mid F(Q)\mid^2\sin^2(z\pi L)$ for $8 \sim 16$ meV , and $\mid F(Q)\mid^2\cos^2(z\pi L)$ for $17 \sim 21$ meV, respectively. Each curve is shifted upward for clarity with the horizontal dashed lines indicating the zero intensity. The vertical dashed lines indicate the non-integral L positions [$L =$ 1.08, 3.24, 5.39 for (a), $L =$ 2.16, 4.31 for (b) ] . (c) and (d) 2D mapping of the spin resonance intensity corresponding to (a) and (b).
 }
\end{figure*}

\begin{figure*}[t]
\renewcommand\thefigure{S7}
\includegraphics[width=0.8\textwidth]{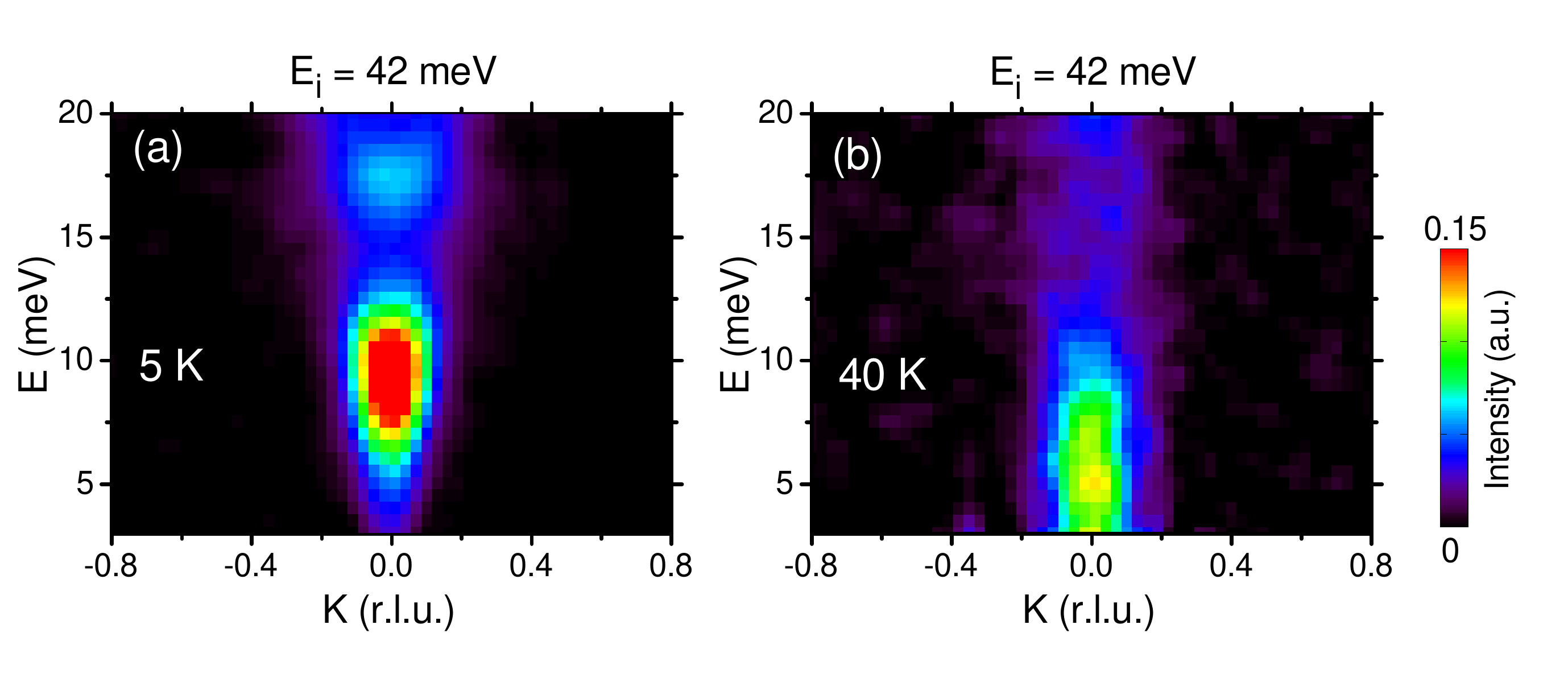}
\caption{ 2D slices of $E$ versus $K$  at $T=$ 5 K and 40 K measured by TOF experiments.
 }
\end{figure*}

\begin{figure*}[t]
\renewcommand\thefigure{S8}
\includegraphics[width=0.69\textwidth]{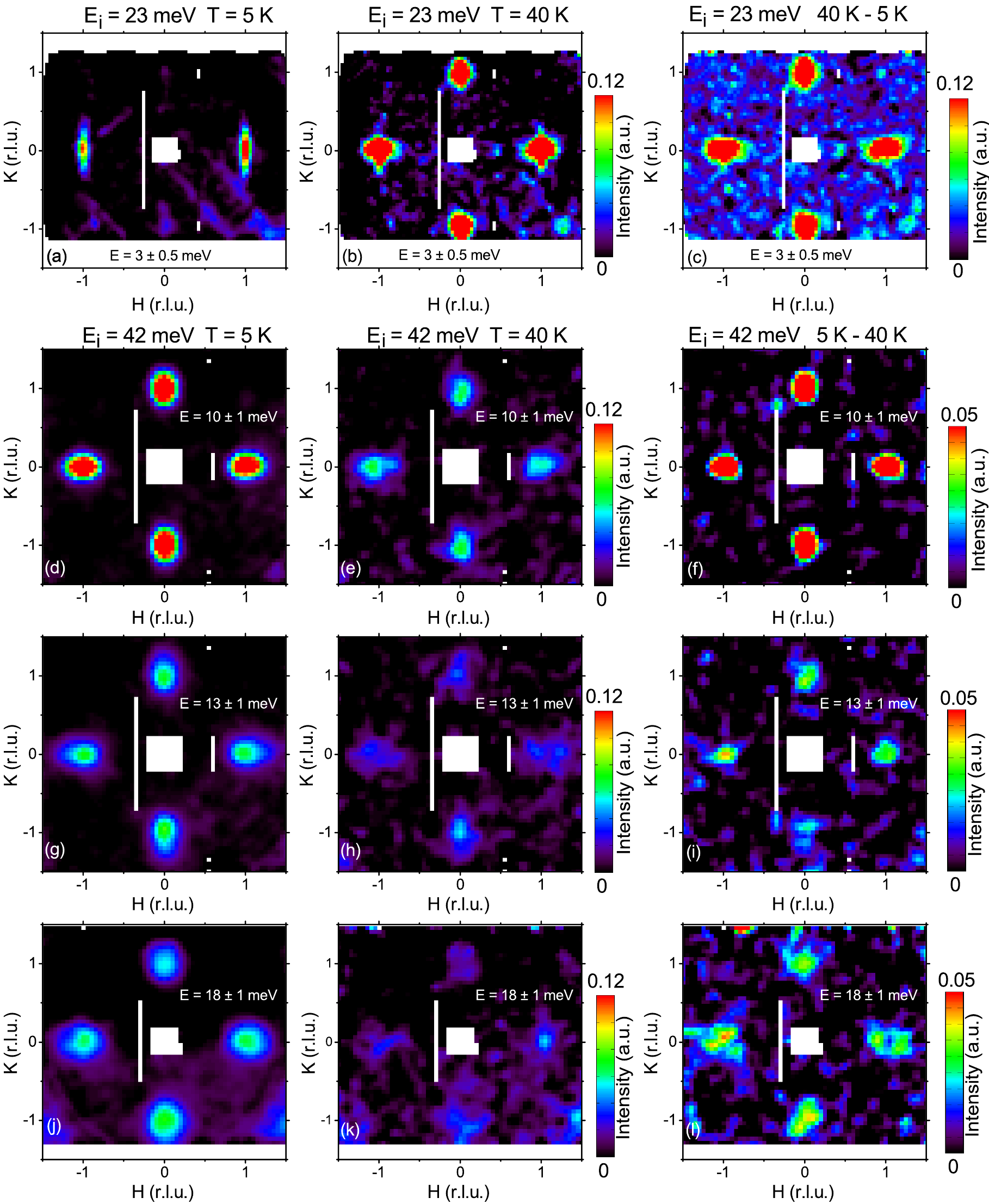}
\caption{
2D constant-energy slices for the spin gap at $E = 3$ meV and three resonance modes at $E = 10,13,18$ meV, for $T=$ 5 K and 40 K, and their differences.
 }
\end{figure*}

\begin{figure*}[t]
\renewcommand\thefigure{S9}
\includegraphics[width=0.69\textwidth]{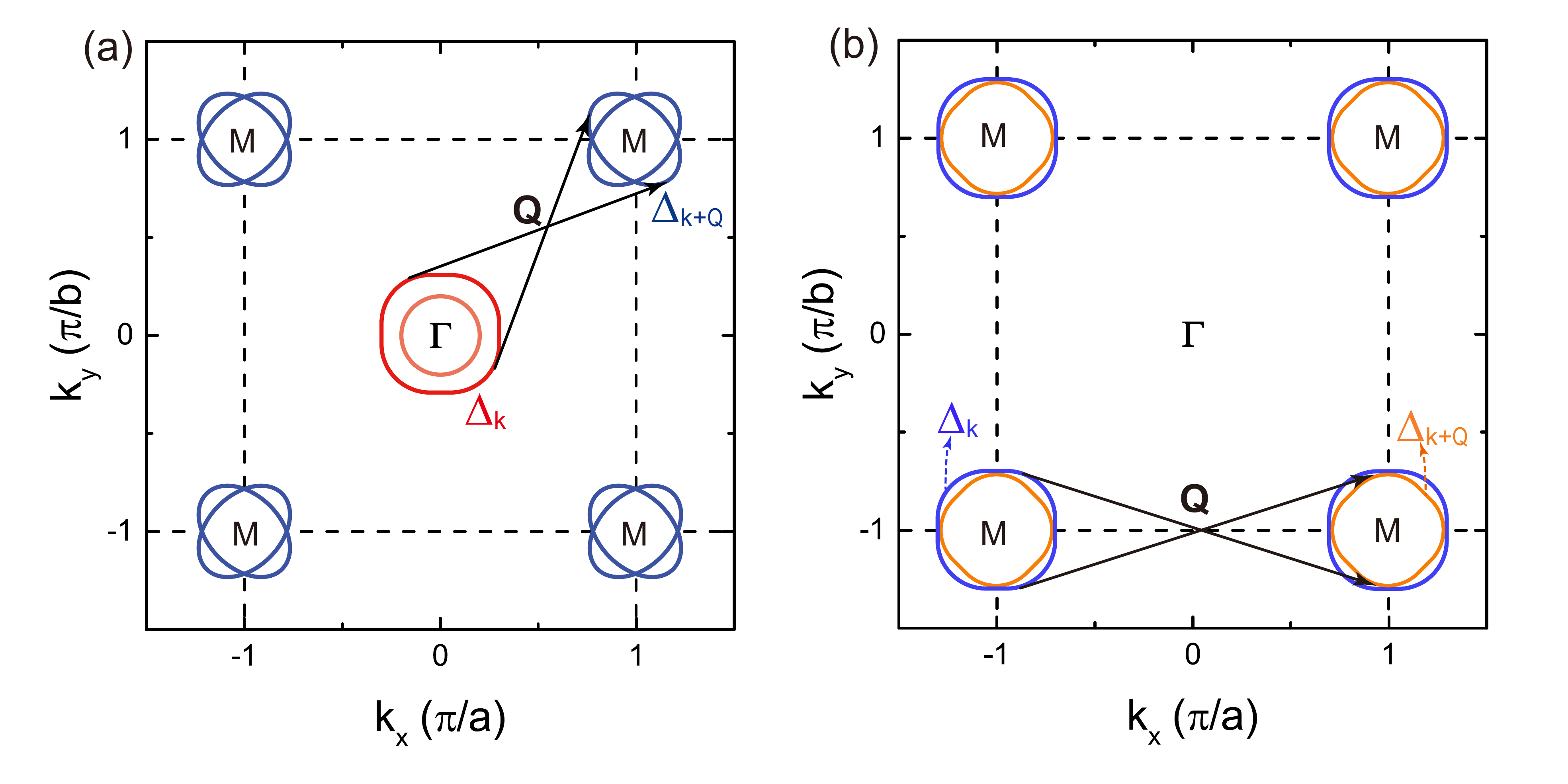}
\caption{
Two cases of Fermi surface nesting in iron pnictides and iron chalcogenides. The sign-reversed superconducting gaps are marked by different colors.
 }
\end{figure*}

\begin{figure*}[t]
\renewcommand\thefigure{S10}
\includegraphics[width=0.8\textwidth]{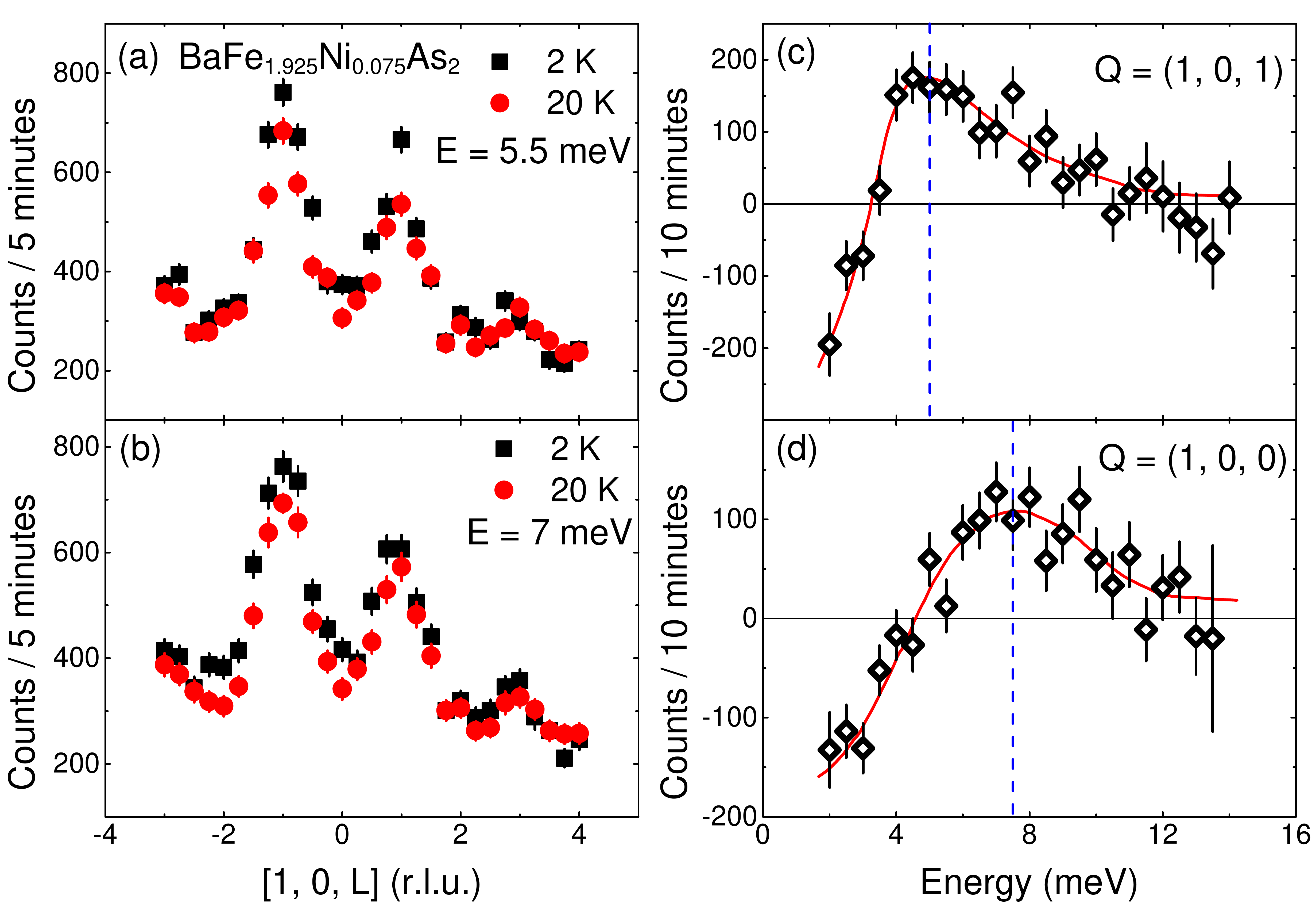}
\caption{
(a)(b) Strong $L-$modulation of spin excitations in underdoped BaFe$_{1.925}$Ni$_{0.075}$As$_2$; (c)(d) Broadening of spin resonance peak and $L-$dependent resonance energy \cite{mywang2010}.
 }
\end{figure*}

\begin{figure*}[t]
\renewcommand\thefigure{S11}
\includegraphics[width=0.8\textwidth]{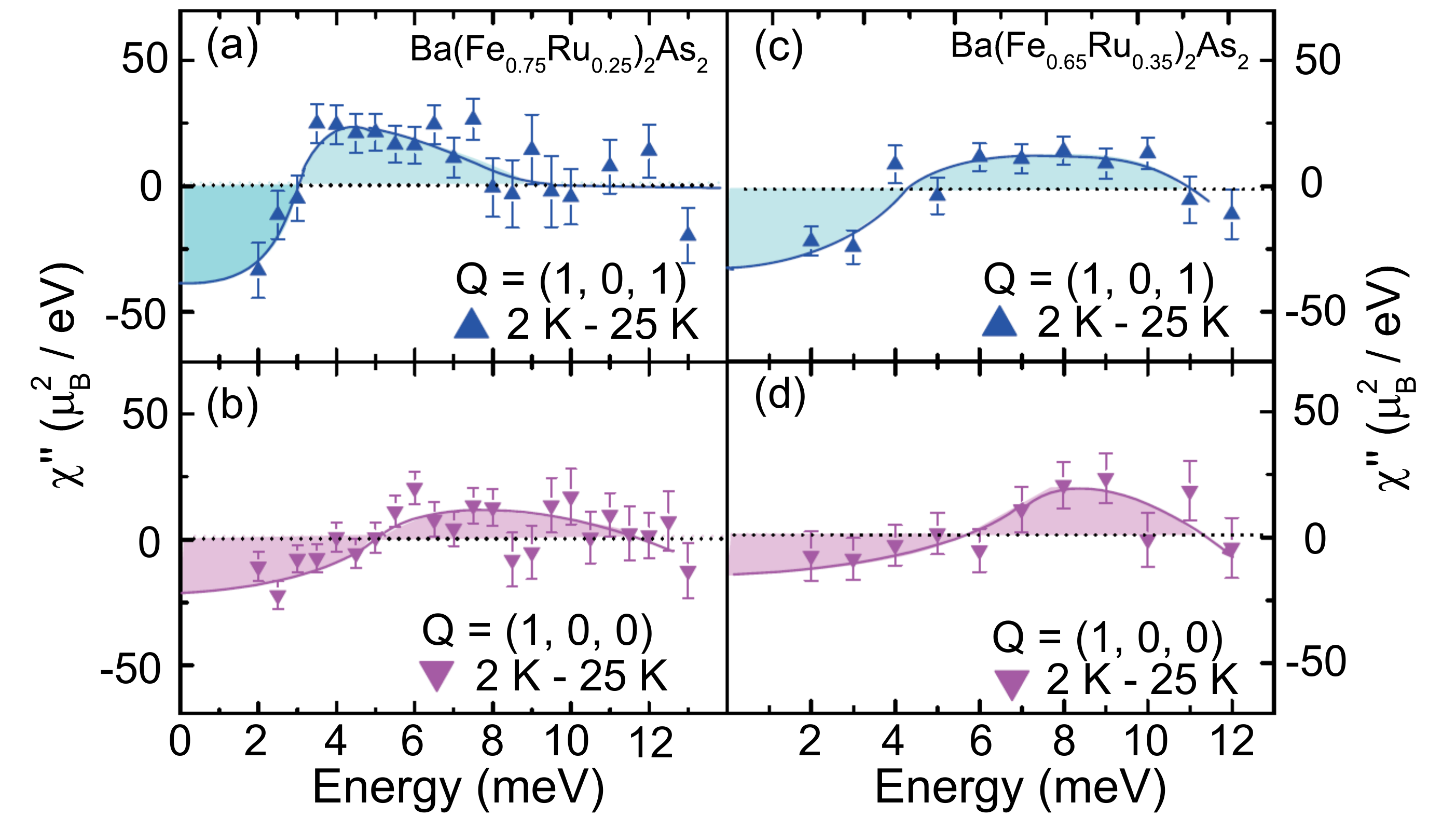}
\caption{
Strong $L-$dependent spin resonance in Ba(Fe$_{1-x}$Ru$_{x}$)$_2$As$_2$ system with $x=0.25$ (underdoped) and $x=0.35$ (optimally doped) \cite{jzhao2013}.
 }
\end{figure*}

\begin{figure*}[t]
\renewcommand\thefigure{S12}
\includegraphics[width=0.75\textwidth]{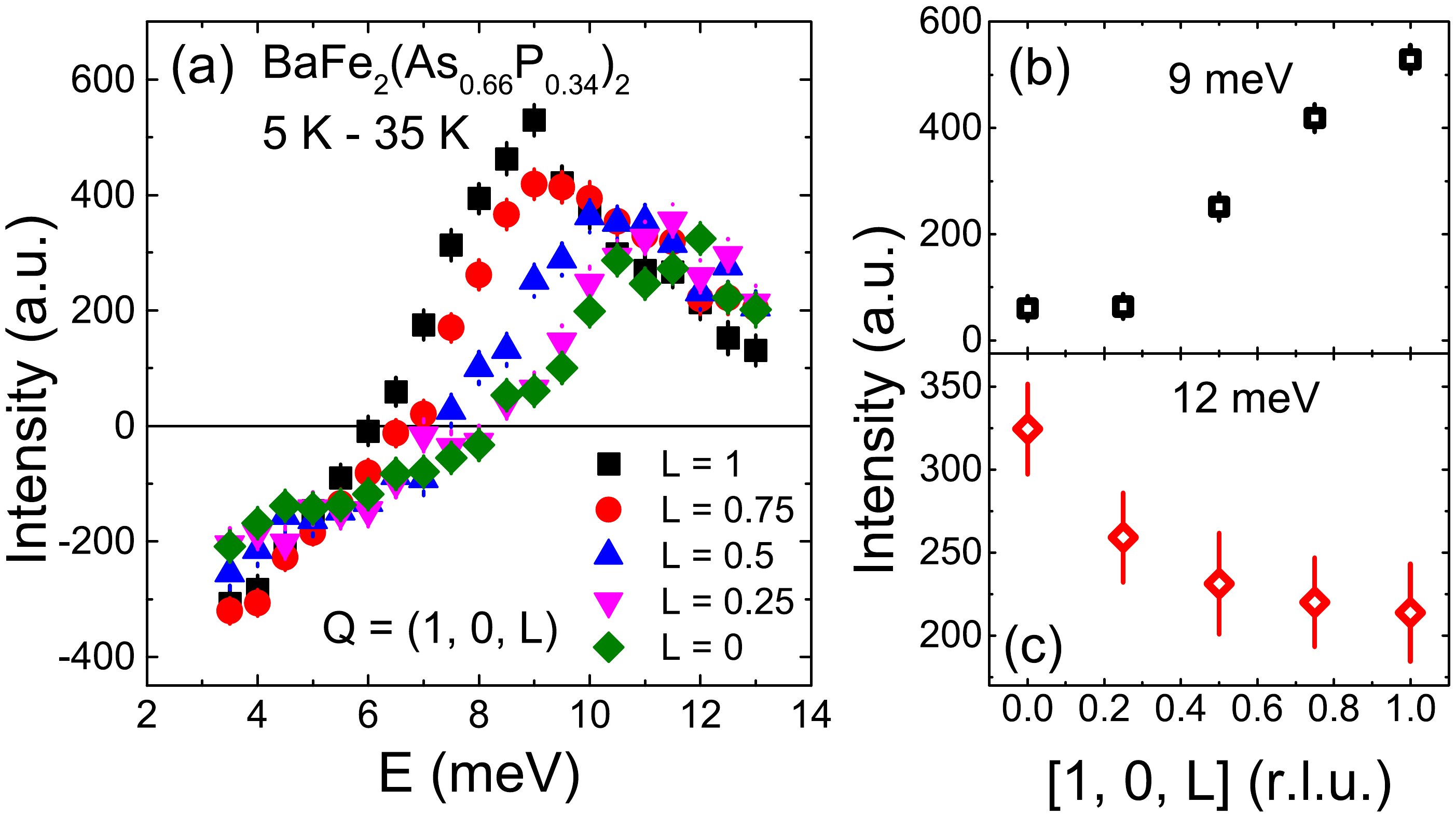}
\caption{
(a)$L-$dispersion of spin resonance peak in optimally doped BaFe$_2$(As$_{1-x}$P$_x$)$_2$. (b)(c) Opposite $L-$dependence of spin resonance intensity at 9 meV and 12 meV \cite{chlee2013}.
 }
\end{figure*}


\begin{thebibliography}{}

\bibitem{jttranquada2014} J. M. Tranquada, G. Xu, and I. A. Zaliznyak, J. Magn. Magn. Mater. {\bf 350}, 148 (2014).
\bibitem{inosov2016} D. S. Inosov, C. R. Phys. {\bf 17}, 60 (2016).
\bibitem{pdai2015} P. Dai, Rev. Mod. Phys. {\bf 87}, 855 (2015).
\bibitem{ostockert2011} O. Stockert, J. Arndt, E. Faulhaber, C. Geibel, H. S. Jeevan, S. Kirchner, M. Loewenhaupt, K. Schmalzl, W. Schmidt, Q. Si, and F. Steglich, Nat.Phys. {\bf 7}, 119 (2011).
\bibitem{meschrig2006}  M. Eschrig, Adv. Phys. {\bf 55}, 47 (2006).
\bibitem{ysidis2007} Y. Sidis, S. Pailh\`{e}s, V. Hinkov, B. Fauqu\'{e}, C. Ulrich, L. Capogna, A. Ivanov, L.-P. Regnault, B. Keimer, and P. Bourges, C. R. Phys. {\bf 8}, 745 (2007).
\bibitem{gyu2009} G. Yu, Y. Li, E. M. Motoyama and M. Greven, Nat. Phys. {\bf 5}, 873 (2009).
\bibitem{si2016} Q. Si, R. Yu, and E. Abrahams, Nat. Rev. Mater. {\bf 1}, 16017(2016).
\bibitem{prichard2011} P. Richard, T. Sato, K. Nakayama, T. Takahashi, and H. Ding, Rep. Prog. Phys. {\bf 74}, 124512 (2011).
\bibitem{korshunov2008} M. M. Korshunov and I. Eremin, Phys. Rev. B {\bf 78}, 140509(R) (2008).
\bibitem{avchubukov2008} A. V. Chubukov, D. V. Efremov, and I. Eremin, Phys. Rev. B {\bf 78}, 134512 (2008).
\bibitem{tamaier2009} T. A. Maier, S. Graser, D. J. Scalapino, and P. Hirschfeld, Phys. Rev. B {\bf 79}, 134520 (2009).
\bibitem{mazin2009} I. Mazin and J. Schmalian, Physica (Amsterdam) {\bf 469C}, 614 (2009).
\bibitem{Seo2008}K.J. Seo, B. A. Bernevig, and J. Hu, Phys. Rev. Lett. {\bf 101}, 206404 (2008).
\bibitem{thanaguri2010} T. Hanaguri, S. Niitaka, K. Kuroki, and H. Takagi, Science {\bf 328}, 474 (2010).
\bibitem{hyang2013} H. Yang, Z. Wang, D. Fang, Q. Deng, Q. Wang, Y. Xiang, Y. Yang, and H. Wen, Nat. Commun. {\bf 4}, 2749 (2013).
\bibitem{aakalenyuk2017} A. A. Kalenyuk, A. Pagliero, E.A. Borodianskyi, A. A. Kordyuk, and V. M. Krasnov, Phys. Rev. Lett. {\bf 120}, 067001 (2018).
\bibitem{zydu2016} Z. Du, X. Yang, H. Lin, D. Fang, G. Du, J. Xing, H. Yang, X. Zhu, and H.-H. Wen, Nat. Commun. {\bf 7}, 10565(2016).
\bibitem{zydu2018} Z. Du, X. Yang, D. Altenfeld, Q. Gu, H. Yang, I. Eremin, P. J. Hirschfeld, I. I. Mazin, H. Lin, X. Zhu, and H. -H. Wen, Nat. Phys. {\bf 14}, 134 (2018).
\bibitem{christianson} A. D. Christianson, E. A. Goremychkin, R. Osborn, S. Rosenkranz, M. D. Lumsden, C. D. Malliakas, I. S. Todorov, H. Claus, D. Y. Chung, M. G. Kanatzidis, R. I. Bewley, and  T. Guidi, Nature (London) {\bf 456}, 930 (2008).
\bibitem{yqiu2009}  Y. Qiu, W. Bao, Y. Zhao, C. Broholm, V. Stanev, Z. Tesanovic, Y. C. Gasparovic, S. Chang, J. Hu, B. Qian, M. Fang, and Z. Mao, Phys. Rev. Lett. {\bf 103}, 067008 (2009).
\bibitem{inosov2010} D. S. Inosov, J. T. Park, P. Bourges, D. L. Sun, Y. Sidis, A. Schneidewind, K. Hradil, D. Haug, C. T. Lin, B. Keimer, and V. Hinkov, Nat.Phys. {\bf 6}, 178 (2010).
\bibitem{qureshi2012} N. Qureshi, P. Steffens, Y. Drees, A. C. Komarek, D. Lamago, Y. Sidis, L. Harnagea, H.-J. Grafe, S. Wurmehl, B. B\"{u}chner, and M. Braden, Phys. Rev. Lett. {\bf 108}, 117001 (2012).
\bibitem{wakimoto} S. Wakimoto,  K. Kodama, M. Ishikado, M. Matsuda, R. Kajimoto, M. Arai, K. Kakurai, F. Esaka, A. Iyo, H. Kito, H. Eisaki, and S. Shamoto, J. Phys. Soc. Jpn. {\bf 79}, 074715 (2010).
\bibitem{zhang2013}	C. Zhang, R. Yu, Y. Su, Y. Song, M. Wang, G. Tan, T. Egami, J. A. Fernandez-Baca, E. Faulhaber, Q. Si, and P. Dai, Phys. Rev. Lett. {\bf 111}, 207002 (2013).
\bibitem{pdjohnson} P. D. Johnson, G. Xu, and W. -G. Yin, {\it Iron-Based Superconductivity}(Springer, New York, 2015), pp 165$-$169.
\bibitem{mywang2010} M. Wang, H. Luo, J. Zhao, C. Zhang, M. Wang, K. Marty, S. Chi, J. W. Lynn, A. Schneidewind, S. Li,and P. Dai, Phys. Rev. B {\bf 81}, 174524 (2010).
\bibitem{hqluo2013b} H. Luo, X. Lu, R. Zhang, M. Wang, E. A. Goremychkin, D. T. Adroja, S. Danilkin, G. Deng, Z. Yamani, and P. Dai, Phys. Rev. B. {\bf 88}, 144516 (2013).
\bibitem{schi2009} S. Chi, A. Schneidewind, J. Zhao, L. W. Harriger, L. Li, Y. Luo, G. Cao, Z. Xu, M. Loewenhaupt, J. Hu, and P. Dai, Phys. Rev. Lett. {\bf 102}, 107006 (2009).
\bibitem{zhang2011} C. Zhang, M. Wang, H. Luo, M. Wang, M. Liu, J. Zhao, D. L. Abernathy, T. A. Maier, K. Marty, M. D. Lumsden, S. Chi, S. Chang, J. A. Rodriguez-Rivera, J. W. Lynn, T. Xiang, J. Hu, and P. Dai, Sci. Rep. {\bf 1}, 115 (2011).
\bibitem{chlee2013} C. H. Lee, P. Steffens, N. Qureshi, M. Nakajima, K. Kihou, A. Iyo, H. Eisaki, and M. Braden, Phys. Rev. Lett. {\bf 111}, 167002 (2013).
\bibitem{jzhao2013} J. Zhao, C. R. Rotundu, K. Marty, M. Matsuda, Y. Zhao, C. Setty, E. Bourret-Courchesne, J. Hu, and R. J. Birgeneau, Phys. Rev. Lett. {\bf 110}, 147003 (2013).
\bibitem{mgkim2013} M. G. Kim, G. S. Tucker, D. K. Pratt, S. Ran, A. Thaler, A. D. Christianson, K. Marty, S. Calder, A. Podlesnyak, S. L. Bud¡¯ko, P. C. Canfield, A. Kreyssig, A. I. Goldman, and R. J. McQueeney, Phys. Rev. Lett. {\bf 110}, 177002 (2013).
\bibitem{zhang2013b} C. Zhang, H.-F. Li, Y. Song, Y. Su, G. Tan, T. Netherton, C. Redding, S. V. Carr, O. Sobolev, A. Schneidewind, E. Faulhaber, L. W. Harriger, S. Li, X. Lu, D. X. Yao, T. Das, A. V. Balatsky, T. Br\"{u}ckel, J. W. Lynn, and P. Dai, Phys. Rev. B {\bf 88}, 064504 (2013).
\bibitem{jtpark2011} J. T. Park, G. Friemel, Y. Li, J. -H. Kim, V. Tsurkan, J. Deisenhofer, H. -A. Krug von Nidda, A. Loidl, A. Ivanov, B. Keimer, and D. S. Inosov, Phys. Rev. Lett. {\bf 107}, 177005 (2011).
\bibitem{gfriemel2012} G. Friemel, W. P. Liu, E. A. Goremychkin, Y. Liu, J. T. Park, O. Sobolev, C. T. Lin, B. Keimer, and D. S. Inosov, Europhys. Lett. {\bf 99}, 67004 (2012).
\bibitem{masurmach2015} M. A. Surmach, F. Br\"{u}ckner, S. Kamusella, R. Sarkar, P. Y. Portnichenko, J. T. Park, G. Ghambashidze, H. Luetkens, P. K. Biswas, W. J. Choi, Y. I. Seo, Y. S. Kwon, H.-H. Klauss, and D. S. Inosov, Phys. Rev. B {\bf 91}, 104515 (2015).
\bibitem{qswang2016a} Q. Wang, Y. Shen, B. Pan, Y. Hao, M. Ma, F. Zhou, P. Steffens, K. Schmalzl, T. R. Forrest, M. Abdel-Hafiez, X. Chen, D. A. Chareev, A. N. Vasiliev, P. Bourges, Y. Sidis, H. Cao, and J. Zhao, Nat. Mater. {\bf 15}, 159 (2016).
\bibitem{mwma2017} M. Ma, L. Wang, P. Bourges, Y. Sidis, S. Danilkin, and Y. Li, Phys. Rev. B {\bf 95}, 100504(R) (2017).
\bibitem{txie2018} T. Xie, D. Gong, H. Ghosh, A. Ghosh, M. Soda, T. Masuda, S. Itoh, F. Bourdarot, L.-P. Regnault, S. Danilkin, S. Li, and H. Luo, Phys. Rev. Lett. {\bf 120}, 137001 (2018).
\bibitem{lwharriger2012} L. W. Harriger, O. J. Lipscombe, C. Zhang, H. Luo, M. Wang, K. Marty, M. D. Lumsden, and P. Dai, Phys. Rev. B {\bf 85}, 054511 (2012).
\bibitem{sonari2010} S. Onari, H. Kontani, and M. Sato, Phys. Rev. B {\bf 81}, 060504(R) (2010).
\bibitem{sonari2012} S. Onari and H. Kontani, Phys. Rev. Lett. {\bf 109}, 137001 (2012).
\bibitem{qswang2016} Q. Wang, J. T. Park, Y. Feng, Y. Shen, Y. Hao, B. Pan, J. W. Lynn, A. Ivanov, S. Chi, M. Matsuda, H. Cao, R. J. Birgeneau, D. V. Efremov, and J. Zhao, Phys. Rev. Lett. {\bf 116}, 197004 (2016).
\bibitem{hqluo2013} H. Luo, M. Wang, C. Zhang, X. Lu, L.-P. Regnault, R. Zhang, S. Li, J. Hu, and P. Dai, Phys. Rev. Lett. {\bf 111}, 107006 (2013).
\bibitem{steffens2013} P. Steffens, C. H. Lee, N. Qureshi, K. Kihou, A. Iyo, H. Eisaki, and M. Braden, Phys. Rev. Lett. {\bf 110}, 137001 (2013).
\bibitem{mwma2017b} M. Ma, P. Bourges, Y. Sidis, Y. Xu, S. Li, B. Hu, J. Li, F. Wang, and Y. Li, Phys. Rev. X {\bf 7}, 021025 (2017).
\bibitem{dhu2017} D. Hu, W. Zhang, Y. Wei, B. Roessli, M. Skoulatos, L.-P. Regnault, G. Chen, Y. Song, H. Luo, S. Li, and P. Dai, Phys. Rev. B {\bf 96}, 180503(R) (2017).
\bibitem{wwang2017} W. Wang, J. T. Park, R. Yu, Y. Li, Y. Song, Z. Zhang, A. Ivanov, J. Kulda, and P. Dai, Phys. Rev. B {\bf 95}, 094519 (2017).
\bibitem{spaihes2003} S. Pailh\`{e}s, Y. Sidis, P. Bourges, C. Ulrich, V. Hinkov, L.-P. Regnault, A. Ivanov, B. Liang, C. Lin, C. Bernhard, and B. Keimer, Phys. Rev. Lett. {\bf 91}, 237002 (2003).
\bibitem{spaihes2004} S. Pailh\`{e}s, Y. Sidis, P. Bourges, V. Hinkov, A. Ivanov, C. Ulrich, L.-P. Regnault, and B. Keimer, Phys. Rev. Lett. {\bf 93}, 167001 (2004).
\bibitem{lcapogna2007} L. Capogna, B. Fauqu\'{e}, Y. Sidis, C. Ulrich, P. Bourges, S. Pailh\`{e}s,  A. Ivanov, J. L. Tallon, B. Liang, C. T. Lin, A. I. Rykov, and B. Keimer, Phys. Rev. B {\bf 75}, 060502(R) (2007).
\bibitem{aiyo2016} A. Iyo, K. Kawashima, T. Kinjo, T. Nishio, S. Ishida, H. Fujihisa, Y. Gotoh, K. Kihou, H. Eisaki, and Y. Yoshida, J. Am. Chem. Soc. {\bf 138}, 3410 (2016).
\bibitem{wrmeier2016} W. R. Meier, T. Kong, U. S. Kaluarachchi, V. Taufour, N. H. Jo, G. Drachuck, A. E. B\"{o}hmer, S. M. Saunders, A. Sapkota, A. Kreyssig, M. A. Tanatar, R. Prozorov, A. I. Goldman, F. F. Balakirev, A. Gurevich, S. L. Bud'ko, and P. C. Canfield, Phys. Rev. B {\bf 94}, 064501 (2016).
\bibitem{wrmeier2017} W. R. Meier, T. Kong, S. L. Bud'ko, and P. C. Canfield, Phys. Rev. Mater. {\bf 1}, 013401 (2017).
\bibitem{supplementary} See Supplemental Material for the sample characterization and raw data of neutron scattering experiments, which includes Refs. [57$-$63].
\bibitem{qfan2015} Q. Fan, W. H. Zhang, X. Liu, Y. J. Yan, M. Q. Ren, R. Peng, H. C. Xu, B. P. Xie, J. P. Hu, T. Zhang, and D. L. Feng, Nat. Phys. {\bf 11},946 (2015).
\bibitem{dfliu2012} D. Liu {\it et al.}, Nat. Commun. {\bf 3}, 931(2012).
\bibitem{slhe2013} S. He {\it et al.}, Nat. Mater. {\bf 12}, 605(2013).
\bibitem{sytan2013} S. Tan, Y. Zhang, M. Xia, Z. Ye, F. Chen, X. Xie, R. Peng, D. Xu, Q. Fan, H. Xu, J. Jiang, T. Zhang, X. Lai, T. Xiang, J. Hu, B. Xie, and Donglai Feng, Nat. Mater. {\bf 12}, 634(2013).
\bibitem{jfhe2014} J. He {\it et al.}, Proc. Natl. Acad. Sci. USA {\bf 111}, 18501 (2014).
\bibitem{lzhao2016} L. Zhao {\it et al.}, Nat. Commun. {\bf 7}, 10608 (2016).
\bibitem{xliu2014} X. Liu {\it et al.}, Nat. Commun. {\bf 5}, 5047(2014).
\bibitem{stuhr} U. Stuhr, B. Roessli, S. Gvasaliya, H. M. R{\o}nnow, U. Filges, D. Graf, A. Bollhalder, D. Hohl , R. B\"{u}rge, M. Schild, L. Holitzner, C. Kaegi, P. Keller, and T. M\"{u}hlebach, Nucl. Instrum. Methods Phys. Res., Sect. A {\bf 853}, 16 (2017).
\bibitem{siki1} M. Nakamura, R. Kajimoto, Y. Inamura, F. Mizuno, M. Fujita, T. Yokoo, and M. Arai, J. Phys. Soc. Jpn. {\bf 78}, 093002 (2009).
\bibitem{Kajimoto2011} R. Kajimoto {\it et al.}, J. Phys. Soc. Jpn. {\bf 80}, SB025 (2011).
\bibitem{Inamura2013} Y. Inamura, T. Nakatani, J. Suzuki, and T. Otomo, J. Phys. Soc. Jpn. {\bf 82}, SA031 (2013).
\bibitem{Meier2018} W. R. Meier, Q.-P. Ding, A. Kreyssig, S. L. Bud¡¯ko, A. Sapkota, K. Kothapalli, V. Borisov, R. Valent¨ª, C. D. Batista, P. P. Orth, R. M. Fernandes, A. I. Goldman, Y. Furukawa, A. E. B\"{o}hmer, and P. C. Canfield, npj Quantum. Mater. {\bf 3}, 5 (2018).
\bibitem{dreznik1996} D. Reznik, P. Bourges, H. F. Fong, L. P. Regnault, J. Bossy, C. Vettier, D. L. Milius, I. A. Aksay, and B. Keimer, Phys. Rev. B {\bf 53}, R14741(R)(1996).
\bibitem{hffong2000} H. F. Fong, P. Bourges, Y. Sidis, L. P. Regnault, J. Bossy, A. Ivanov, D. L. Milius, I. A. Aksay, and B. Keimer, Phys. Rev. B {\bf 61}, 14773 (2000).
\bibitem{flochner2017} F. Lochner, F. Ahn, T. Hickel, and Ilya Eremin, Phys. Rev. B {\bf 96}, 094521 (2017).
\bibitem{dmou2016} D. Mou, T. Kong, W. R. Meier, F. Lochner, L. L. Wang, Q. Lin, Y. Wu, S. L. Bud¡¯ko, I. Eremin, D. D. Johnson, P. C. Canfield, and A. Kaminski, Phys. Rev. Lett. {\bf 117}, 277001 (2016).
\bibitem{nakayama2009} K. Nakayama, T. Sato, P. Richard, Y.-M. Xu, Y. Sekiba, S. Souma, G. F. Chen, J. L. Luo, N. L. Wang, H. Ding, and T. Takahashi, Europhys. Lett. {\bf 85}, 67002 (2009).
\bibitem{nakayama2011} K. Nakayama, T. Sato, P. Richard, Y.-M. Xu, T. Kawahara, K. Umezawa, T. Qian, M. Neupane, G. F. Chen, H. Ding, and T. Takahashi, Phys. Rev. B {\bf 83}, 020501(R) (2011).
\bibitem{chlee2016} C. H. Lee, K. Kihou, J. T. Park, K. Horigane, K. Fujita, F. Wa{\ss}er, N. Qureshi, Y. Sidis, J. Akimitsu, and M. Braden, Sci. Rep. {\bf 6}, 23424 (2016).
\bibitem{kterashima2009} K. Terashima, Y. Sekiba, J. H. Bowen, K. Nakayama, T. Kawahara, T. Sato, P. Richard, Y.-M. Xu, L. J. Li, G. H. Cao, Z.-A. Xu, H. Ding, and T. Takahashi, Proc. Natl. Acad. Sci. USA {\bf 106}, 7330 (2009).
\bibitem{mwang2016} M. Wang, M. Yi, H. L. Sun, P. Valdivia, M. G. Kim, Z. J. Xu, T. Berlijn, A. D. Christianson, S. Chi, M. Hashimoto, D. H. Lu, X. D. Li, E. Bourret-Courchesne, P. Dai, D. H. Lee, T. A. Maier, and R. J. Birgeneau, Phys. Rev. B {\bf 93}, 205149 (2016).
\bibitem{yzhang2012} Y. Zhang, Z. Ye, Q. Ge, F. Chen, J. Jiang, M. Xu, B. Xie, and D. Feng, Nat. Phys. {\bf 8}, 371 (2012).
\bibitem{jmaletz2014} J. Maletz, V. B. Zabolotnyy, D. V. Evtushinsky, S. Thirupathaiah, A. U. B. Wolter, L. Harnagea, A. N. Yaresko, A. N. Vasiliev, D. A. Chareev, A. E. B\"{o}hmer, F. Hardy, T. Wolf, C. Meingast, E. D. L. Rienks, B. B\"{u}chner, and S. V. Borisenko, Phys. Rev. B {\bf 89}, 220506(R)(2014).
\bibitem{hmiao2012} H. Miao, P. Richard, Y. Tanaka, K. Nakayama, T. Qian, K. Umezawa, T. Sato, Y.-M. Xu, Y. B. Shi, N. Xu, X.-P. Wang, P. Zhang, H.-B. Yang, Z.-J. Xu, J. S. Wen, G.-D. Gu, X. Dai, J.-P. Hu, T. Takahashi, and H. Ding, Phys. Rev. B {\bf 85}, 094506 (2012).
\bibitem{qqge2013} Q. Q. Ge, Z. R. Ye, M. Xu, Y. Zhang, J. Jiang, B. P. Xie, Y. Song, C. L. Zhang, P. Dai, and D. L. Feng, Phys. Rev. X {\bf 3}, 011020 (2013).
\bibitem{zswang2012} Z.-S. Wang, Z.-Y. Wang, H.-Q. Luo, X.-Y. Lu, J. Zhu, C.-H. Li, L. Shan, H. Yang, H.-H. Wen, and C. Ren, Phys. Rev. B {\bf 86}, 060508(R) (2012).
\bibitem{pustovit2016} Y. V. Pustovit, and A. A. Kordyuk, J. Low Temp. Phys. {\bf 42}, 995 (2016).
\bibitem{jzhu2015} J. Zhu, Z. Wang, Z. Wang, X. Hou, H. Luo, X. Lu, C. Li, L. Shan, H. Wen, and C. Ren, Chin. Phys. Lett. {\bf 32}, 077401 (2015).
\bibitem{yzhang2011} Y. Zhang, L. X. Yang, M. Xu, Z. R. Ye, F. Chen, C. He, H. C. Xu, J. Jiang, B. P. Xie, J. J. Ying, X. F. Wang, X. H. Chen, J. P. Hu, M. Matsunami, S. Kimura, and D. L. Feng, Nat. Mater. {\bf 10}, 273 (2011).
\bibitem{xhniu2016} X. H. Niu, S. D. Chen, J. Jiang, Z. R. Ye, T. L. Yu, D. F. Xu, M. Xu, Y. Feng, Y. J. Yan, B. P. Xie, J. Zhao, D. C. Gu, L. L. Sun, Q. Mao, H. Wang, M. Fang, C. J. Zhang, J. P. Hu, Z. Sun, and D. L. Feng, Phys. Rev. B {\bf 93}, 054516 (2016).
\bibitem{kiida2017} K. Iida, M. Ishikado, Y. Nagai, H. Yoshida, A. D. Christianson, N. Murai, K. Kawashima, Y. Yoshida, H. Eisaki, and A. Iyo, J. Phys. Soc. Jpn. {\bf 86}, 093703 (2017).
\bibitem{ryang2017} R. Yang, Y. Dai, B. Xu, W. Zhang, Z. Qiu, Q. Sui, C. C. Homes, and X. Qiu, Phys. Rev. B {\bf 95}, 064506 (2017).
\bibitem{pkbiswas2016} P. K. Biswas, A. Iyo, Y. Yoshida, H. Eisaki, K. Kawashima, and A. D. Hillier, Phys. Rev. B {\bf 95}, 140505(R) (2017).
\bibitem{kcho2017} K. Cho, A. Fente, S. Teknowijoyo, M. A. Tanatar, K. R. Joshi, N. M. Nusran, T. Kong, W. R. Meier, U. Kaluarachchi, I. Guillam\'{o}n, H. Suderow, S. L. Bud'ko, P. C. Canfield, and R. Prozorov, Phys. Rev. B {\bf 95}, 100502(R) (2017).
\bibitem{jcui2017} J. Cui, Q.-P. Ding, W. R. Meier, A. E. B\"{o}hmer, T. Kong, V. Borisov, Y. Lee, S. L. Bud'ko, R. Valent\'{\i}, P. C. Canfield, and Y. Furukawa, Phys. Rev. B {\bf 96}, 104512 (2017).
\bibitem{qpding2017} Q. -P. Ding, W. R. Meier, A. E. B\"{o}hmer, S. L. Bud'ko, P. C. Canfield, and Y. Furukawa, Phys. Rev. B {\bf 96}, 220510(R) (2017).
\bibitem{jzhao2012} J. Zhao, D. T. Adroja, D. -X. Yao, R. Bewley, S. Li, X. F.Wang, G. Wu, X. H. Chen, J. Hu, and P. Dai, Nat. Phys. {\bf 5}, 555 (2009).
\bibitem{jtpark2012} J. T. Park, G. Friemel, T. Loew, V. Hinkov, Y. Li, B. H. Min, D. L. Sun, A. Ivanov, A. Piovano, C. T. Lin, B. Keimer, Y. S. Kwon, and D. S. Inosov, Phys. Rev. B {\bf 86}, 024437 (2012).
\bibitem{chlee2016b} K. Horigane, K. Kihou, K. Fujita, R. Kajimoto, K. Ikeuchi, S. Ji, J. Akimitsu and C. H. Lee,  Sci. Rep. {\bf 6}, 33303 (2016).
\bibitem{mwang2013} M. Wang, C. Zhang, X. Lu, G. Tan, H. Luo, Y. Song, M. Wang, X. Zhang, E.A. Goremychkin, T.G. Perring, T.A. Maier, Z. Yin, K. Haule, G. Kotliar and P. Dai, Nat. Commun. {\bf 4}, 2874(2013).
\bibitem{nni2008} N. Ni, S. Nandi, A. Kreyssig, A. I. Goldman, E. D. Mun, S. L. Bud'ko, and P. C. Canfield, Phys. Rev. B {\bf 78}, 014523 (2008).
\bibitem{mrotter2008} M. Rotter, M. Tegel, D. Johrendt, I. Schellenberg, W. Hermes, and R. P\"{o}ttgen, Phys. Rev. B {\bf 78}, 020503(R) (2008).
\bibitem{hqluo2008} H. Luo, Z. Wang, H. Yang, P. Cheng, X. Zhu, and H. -H. Wen, Supercond. Sci. Technol. {\bf 21}, 125014 (2008).
\bibitem{kkihou2010} K. Kihou, T. Saito, S. Ishida, M. Nakajima, Y. Tomioka, H. Fukazawa, Y. Kohori, T. Ito, S. Uchida, A. Iyo, C.-H. Lee, and H. Eisaki, J. Phys. Soc. Jpn. {\bf 79}, 124713 (2010).

\end{thebibliography}

\begin{thebibliography}{}
\bibitem{Meier2016} W. R. Meier, T. Kong, U. S. Kaluarachchi, V. Taufour, N. H. Jo, G. Drachuck, A. E. B\"{o}hmer, S. M. Saunders, A. Sapkota, A. Kreyssig, M. A. Tanatar, R. Prozorov, A. I. Goldman, F. F. Balakirev, A. Gurevich, S. L. Bud'ko, and P. C. Canfield, Phys. Rev. B {\bf 94}, 064501 (2016).
\bibitem{Meier2017} W. R. Meier, T. Kong, S. L. Bud¡¯ko, and P. C. Canfield, Phys. Rev. Materials {\bf 1}, 013401 (2017).
\bibitem{aiyo2016}  A. Iyo, K. Kawashima, T. Kinjo, T. Nishio, S. Ishida, H. Fujihisa, Y. Gotoh, K. Kihou, H. Eisaki, and Y. Yoshida, J. Am. Chem. Soc. {\bf 138}, 3410 (2016).
\bibitem{dmou2016} D. Mou, T. Kong, W. R. Meier, F. Lochner, L. Wang, Q. Lin, Y. Wu, S. L. Bud¡¯ko, I. Eremin, D. D. Johnson, P. C. Canfield, and A. Kaminski, Phys. Rev. Lett. {\bf 117}, 277001 (2016).
\bibitem{flochner2017} F. Lochner, F. Ahn, T. Hickel, and Ilya Eremin, Phys. Rev. B {\bf 96}, 094521 (2017).
\bibitem{Meier2018} W. R. Meier, Q.-P. Ding, A. Kreyssig, S. L. Bud¡¯ko, A. Sapkota, K. Kothapalli, V. Borisov, R. Valent¨ª, C. D. Batista, P. P. Orth, R. M. Fernandes, A. I. Goldman, Y. Furukawa, A. E. B\"{o}hmer, and P. C. Canfield, npj Quan. Mat. {\bf 3}, 5 (2018).
\bibitem{Pailhes1} S. Pailh\`{e}s, Y. Sidis, P. Bourges, C. Ulrich, V. Hinkov, L.-P. Regnault, A. Ivanov, B. Liang, C. Lin, C. Bernhard, and B. Keimer, Phys. Rev. Lett. {\bf 91}, 237002 (2003).
\bibitem{Pailhes2} S. Pailh\`{e}s, Y. Sidis, P. Bourges, V. Hinkov, A. Ivanov, C. Ulrich, L.-P. Regnault, and B. Keimer, Phys. Rev. Lett. {\bf 93}, 167001 (2004).
\bibitem{Kajimoto2011} R. Kajimoto, M. Nakamura, Y. Inamura, F. Mizuno, K. Nakajima, S. Ohira-Kawamura, T. Yokoo, T. Nakatani, R. Maruyama, K. Soyama, K. Shibata, K. Suzuya, S. Sato, K. Aizawa, M. Arai, S. Wakimoto, M. Ishikado, S. Shamoto, M. Fujita, H. Hiraka, K. Ohoyama, K. Yamada, and C.-H. Lee, J. Phys. Soc. Jpn. {\bf 80}, SB025 (2011).
\bibitem{Inamura2013} Y. Inamura, T. Nakatani, J. Suzuki, and T. Otomo, J. Phys. Soc. Jpn. {\bf 82}, SA031 (2013).
\bibitem{gyu2009} G. Yu, Y. Li, E. M. Motoyama and M. Greven, Nat. Phys. {\bf 5}, 873 (2009).
\bibitem{korshunov2008} M. M. Korshunov and I. Eremin, Phys. Rev. B {\bf 78}, 140509(R) (2008).
\bibitem{avchubukov2008} A. V. Chubukov, D. V. Efremov, and I. Eremin, Phys. Rev. B {\bf 78}, 134512 (2008).
\bibitem{mazin2009} I. Mazin and J. Schmalian, Physica C {\bf 469}, 614 (2009).
\bibitem{tamaier2009} T. A. Maier, S. Graser, D. J. Scalapino, and P. Hirschfeld, Phys. Rev. B {\bf 79}, 134520 (2009).
\bibitem{qswang2016} Q. Wang, J. T. Park, Y. Feng, Y. Shen, Y. Hao, B. Pan, J. W. Lynn, A. Ivanov, S. Chi, M. Matsuda, H. Cao, R. J. Birgeneau, D. V. Efremov, and J. Zhao, Phys. Rev. Lett. {\bf 116}, 197004 (2016).
\bibitem{sonari2010} S. Onari, H. Kontani, and M. Sato, Phys. Rev. B {\bf 81}, 060504(R) (2010).
\bibitem{sonari2012} S. Onari and H. Kontani, Phys. Rev. Lett. {\bf 109}, 137001 (2012).
\bibitem{qfan2015} Q. Fan, W. H. Zhang, X. Liu, Y. J. Yan, M. Q. Ren, R. Peng, H. C. Xu, B. P. Xie, J. P. Hu, T. Zhang, and D. L. Feng,Nat. Phys. {\bf 11},946 (2015).
\bibitem{dfliu2012} D. Liu, W. Zhang, D. Mou, J. He, Y.-B. Ou, Q.-Y. Wang, Z. Li, L. Wang, L. Zhao, S. He, Y. Peng, X. Liu, C. Chen, L. Yu, G. Liu, X. Dong, J. Zhang, C. Chen, Z. Xu, J. Hu, X. Chen, X. Ma, Q. Xue, and X.J. Zhou, Nat. Commun. {\bf 3}, 931(2012).
\bibitem{slhe2013} S. He, J. He, W. Zhang, L. Zhao, D. Liu, X. Liu, D. Mou, Y.-B. Ou, Q.-Y. Wang, Z. Li, L. Wang, Y. Peng, Y. Liu, C. Chen, L. Yu, G. Liu, X. Dong, J. Zhang, C. Chen, Z. Xu, X. Chen, X. Ma, Q. Xue, and X. J. Zhou, Nat. Mater. {\bf 12}, 605(2013).
\bibitem{sytan2013} S. Tan, Y. Zhang, M. Xia, Z. Ye, F. Chen, X. Xie, R. Peng, D. Xu, Q. Fan, H. Xu, J. Jiang, T. Zhang, X. Lai, T. Xiang, J. Hu, B. Xie, and Donglai Feng, Nat. Mater. {\bf 12}, 634(2013).
\bibitem{jfhe2014} J. He, X. Liu, W. Zhang, L. Zhao, D. Liu, S. He, D. Mou, F. Li, C. Tang, Z. Li, L. Wang, Y. Peng, Y. Liu, C. Chen, L. Yu, G. Liu, X. Dong, J. Zhang, C. Chen, Z. Xu, X. Chen, X. Ma, Q. Xue, and X. J. Zhou, Proc. Natl. Acad. Sci. USA {\bf 111}, 18501 (2014).
\bibitem{lzhao2016} L. Zhao, A. Liang, D. Yuan, Y. Hu, D. Liu, J. Huang, S. He, B. Shen, Y. Xu, X. Liu, L. Yu, G. Liu, H. Zhou, Y. Huang, X. Dong, F. Zhou, K. Liu, Z. Lu, Z. Zhao, C. Chen, Z. Xu, and X. J. Zhou, Nat. Commun. {\bf 7}, 10608 (2016).
\bibitem{xliu2014} X. Liu, D. Liu, W. Zhang, J. He, L. Zhao, S. He, D. Mou, F. Li, C. Tang, Z. Li, L. Wang, Y. Peng, Y. Liu, C. Chen, L. Yu, G. Liu, X. Dong, J. Zhang, C. Chen, Z. Xu, X. Chen, X. Ma, Q. Xue, and X. J. Zhou, Nat. Commun. {\bf 5}, 5047(2014).
\bibitem{jtpark2011} J. T. Park, G. Friemel, Y. Li, J. -H. Kim, V. Tsurkan, J. Deisenhofer, H. -A. Krug von Nidda, A. Loidl, A. Ivanov, B. Keimer, and D. S. Inosov, Phys. Rev. Lett. {\bf 107}, 177005 (2011).
\bibitem{gfriemel2012} G. Friemel, W. P. Liu, E. A. Goremychkin, Y. Liu, J. T. Park, O. Sobolev, C. T. Lin, B. Keimer, and D. S. Inosov, Europhys. Lett. {\bf 99}, 67004 (2012).
\bibitem{zydu2016} Z. Du, X. Yang, H. Lin, D. Fang, G. Du, J. Xing, H. Yang, X. Zhu, and H.-H. Wen, Nat. Commun. {\bf 7}, 10565(2016).
\bibitem{zydu2018} Z. Du, X. Yang, D. Altenfeld, Q. Gu, H. Yang, I. Eremin, P. J. Hirschfeld, I. I. Mazin, H. Lin, X. Zhu and H. -H. Wen, Nat. Phys. {\bf 14}, 134 (2018).
\bibitem{ryang2017} R. Yang, Y. Dai, B. Xu, W. Zhang, Z. Qiu, Q. Sui, C. C. Homes, and X. Qiu, Phys. Rev. B {\bf 95}, 064506 (2017).
\bibitem{pkbiswas2016} P. K. Biswas, A. Iyo, Y. Yoshida, H. Eisaki, K. Kawashima, A. D. Hillier, Phys. Rev. B {\bf 95}, 140505(R) (2017).
\bibitem{kcho2017} K. Cho, A. Fente, S. Teknowijoyo, M. A. Tanatar, K. R. Joshi, N. M. Nusran, T. Kong, W. R. Meier, U. Kaluarachchi, I. Guillam\'{o}n, H. Suderow, S. L. Bud'ko, P. C. Canfield, and R. Prozorov, Phys. Rev. B {\bf 95}, 100502(R) (2017).
\bibitem{dreznik1996} D. Reznik, P. Bourges, H. F. Fong, L. P. Regnault, J. Bossy, C. Vettier, D. L. Milius, I. A. Aksay, and B. Keimer, Phys. Rev. B {\bf 53}, R14741(R)(1996).
\bibitem{pdai2015} P. Dai, Rev. Mod. Phys. {\bf 87}, 855 (2015).
\bibitem{hffong2000} H. F. Fong, P. Bourges, Y. Sidis, L. P. Regnault, J. Bossy, A. Ivanov, D. L. Milius, I. A. Aksay, and B. Keimer, Phys. Rev. B {\bf 61}, 14773 (2000).
\bibitem{inosov2016} D. S. Inosov, C. R. Physique {\bf 17}, 60 (2016).
\bibitem{pdjohnson} P. D. Johnson, G. Xu, W. -G. Yin {\it Iron-Based Superconductivity}, Springer, P165-P169 (2015).
\bibitem{mywang2010} M. Wang, H. Luo, J. Zhao, C. Zhang, M. Wang, K. Marty, S. Chi, J. W. Lynn, A. Schneidewind, S. Li,and P. Dai, Phys. Rev. B {\bf 81}, 174524 (2010).
\bibitem{jzhao2013} J. Zhao, C. R. Rotundu, K. Marty, M. Matsuda, Y. Zhao, C. Setty, E. Bourret-Courchesne, J. Hu, and R. J. Birgeneau, Phys. Rev. Lett. {\bf 110}, 147003 (2013).
\bibitem{chlee2013} C. H. Lee, P. Steffens, N. Qureshi, M. Nakajima, K. Kihou, A. Iyo, H. Eisaki, and M. Braden, Phys. Rev. Lett. {\bf 111}, 167002 (2013).



\end{thebibliography}
\end{document}